\documentclass[10pt]{article}    
\usepackage{amssymb,latexsym,amsmath} 
\usepackage[mathscr]{eucal}  
\newfont{\sffl}{msbm10 at 16pt} 
\newfont{\sff}{msbm10 at 10pt}

\begin{document}           

\title{A closed-form energy-minimization basis for\\
       gravity field source estimation: DIDACKS IV\thanks{\small{Approved for public  release; distribution is unlimited.}}}  
 
\author{Alan Rufty}         
\date{November 28, 2007}

\maketitle                 

\newcommand{\KD}{K_{\text{D}}}
\begin{abstract}
Previous articles in this series presented a (weighted) field energy (i.e., Dirichlet integral) based approach to finding point source solutions to Laplace's equation over specific bounded and unbounded domains, where the sources are assumed to be in the complimentary region.  The associated mathematical framework was labeled a Dirichlet-integral dual-access collocation-kernel space (DIDACKS).  Specifically, for $\mbox{\sff R}^3$ half-space and the exterior of an $\mbox{\sff R}^3$ sphere, which are the primary settings used in geoexploration and physical geodesy, the DIDACKS approach yields exact closed-form linear equation sets for the strengths of point sources when their locations are fixed.  By building on the field energy minimization underpinnings of DIDACKS theory and by making certain natural assumptions about the general nature of the energy/density configuration of the Earth's interior it is shown that the problem of estimating the Earth's interior density, either globally or locally, can be naturally reframed as a energy minimization one.  Although there are certain conceptual complications to be factored in, the basic idea is that a static stable density configuration is a minimum energy configuration, which tends to be unique (when all other things are equal); hence, a field energy minimization approach can be counted on to  generally lead to a physically motivated unique solution.  Techniques touched on here should provide practical implementation tools, or at least some helpful hints, for handling many of the well-known ill-posedness issues associated with mass density estimation and other related inverse-source Laplacian problems.  These and additional associated considerations directly lead to the possibility of flexible and powerful implementations of (local) mass density estimation software that incorporates and extends certain long accepted and commonly used (geoexploration) implementations.  Clearly, the presented techniques can also be directly adapted for use in other areas of applied mathematics as well as other physical problem areas, such as electrostatics.  This article focuses more on overall implementation issues than on concrete specific examples and contains no numerical examples; however, due consideration has been given to the potential viability of the suggested approaches. 
\end{abstract}

\vskip .05in
\noindent
\begin{itemize}
\item[\ \ ] \small{\textbf{Key words:} {Laplace's equation, inverse problem, Dirichlet form, point collocation, reproducing\\ \phantom{Key words. L} kernels,} 
fundamental solutions, point sources, potential theory}
\item[\ \ ] \small{\textbf{AMS subject classification (2000):} {Primary 86A20. Secondary 35J05, 65R99, 86A22}}
\end{itemize}

\newcommand{\SubSec}[1]{

\vskip .18in
\noindent
\underline{{#1}}
\vskip .08in

}

\newcommand{\ls}{\vphantom{\big)}} 
\newcommand{\lsm}{\!\vphantom{\big)}} 

\newcommand{\eq}{:=}
\newcommand{\mLarge}[1]{\text{\begin{Large} $#1$ \end{Large}}}
\newcommand{\mSmall}[1]{\text{\begin{footnotesize} $#1$ \end{footnotesize}}}

\newcommand{\Blbrac}{\rlap{\bigg{\lceil}}\bigg{\lfloor}} 
\newcommand{\Brbrac}{\bigg{\rbrack}} 

\newcommand{\smallindex}[1]{\text{\raisebox {1.5pt} {${}_{#1}$}}}
\newcommand{\smallindexes}[1]{\text{\raisebox {1.0pt} {${}_{\!\hspace{0.5pt} #1}$}}}

\newcommand{\R}[1]{${\mbox{\sff R}}^{#1}$} 
\newcommand{\mR}[1]{{\mbox{\sff R}}^{#1}}
\newcommand{\RR}{${\mbox{\sff R}}$}
\newcommand{\mRR}{{\mbox{\sff R}}}
\newcommand{\C}{${\mbox{\sff C}}$}
\newcommand{\mC}{{\mbox{\sff C}}}
\newcommand{\riii}{$\mbox{\sff R}^3$} 
\newcommand{\mriii}{\mbox{\sff R}^3}

\newcommand{\Dr}{\mathscr{D}_r}

\newcommand{\sps}{0} 
\newcommand{\spss}{0}

\renewcommand {\baselinestretch}{1.35} 

\newpage
\noindent
 \begin{Large}{\textbf{(i)\ \ \ Preamble}}\end{Large} 

\vskip .18in
\noindent
  This article discusses Dirichlet-Integral Based Dual-Access Collocation-Kernel (DIDACKS) based techniques for geoexploration or Laplacian inverse source theory.  Most of the same techniques can (and have) been used for gravity modeling. 

 From a first-cut intuitive perspective, given that physical systems in static equilibrium (or even ones in quasi-static equilibrium like the Earth, which, for simplicity of exposition, is treated here as a static time-independent case) tend to be in minimum energy configurations, and given that DIDACKS fits minimize the (weighted) energy of the entire external field exactly (as well as that of the exterior energy of the error field) and, moreover, that DIDACKS theory was first explicitly developed and tested for the two standard geometries that are the most commonly used ones in geoexploration and physical geodesy---namely the exterior of an $\mbox{\sff R}^3$ sphere and $\mbox{\sff R}^3$ half-space---one might tentatively conclude that in order to solve a geophysical inverse density estimation problem all that is necessary to do is to simply perform a DIDACKS point mass fit and then reinterpret the results.  Here, of course, some sort of natural accommodations to the innate ill-conditioning of the problem must be made by, say, carefully choosing the source spacing and depths.  On second glance, it might appear that this first take is entirely too nieve because the mathematical form of the gravitational field energy itself is negative (that is to say, it is proportional to the negative of the Dirichlet integral of the underlying potential) and thus for gravitational source estimation problems a negative energy solution would require minimizing the negative of the DIDACKS cost function, which would lead to negative run-away solutions that correspond to a worst fit rather than a best fit.  It would thus seem that while DIDACKS theory may be directly applied as it stands to electrostatics or other problems where the energy density is positive and it may be applied to gravitational modeling problems (as discussed at length in \cite{DIDACKS}), it should not be applied to gravitational source problems without some sort of significant modifications.  On further examination, one might be inclined to think that a direct energy minimization approach to gravity source estimation is more or less hopeless since mass density estimation problems tend to be, by their very nature, ill-conditioned, and any nostrum that patches the theory must surely be as likely to hurt as it is to improve the condition number; conversely, due to natural trade-offs that one might expect to have to make between theory and practice, anything done to improve the condition number must, it might seem, surely blemish the theory, in some sense or other.

  Surprisingly, upon considering several simple ideas and their implications, it turns out that the first more optimistic take on things is much closer to the final truth of the matter.  One central point is that it is necessary, at least implicitly, to consider conceptual issues involving the linkage of the internal energy of the source density and the energy content of the exterior field.  When this is done, it turns out that even the first-cut perspective mentioned above is, if anything, actually too pessimistic.   For example, it turns out, upon closer examination, that all of the various relevant standard regularization techniques have a sound theoretical justification and, conversely, all of the considered theoretical nostrums lead to solutions which improve the condition number.  When one steps back to consider what the implications of all of this might be for the application of the DIDACKS approach to other arenas, there is even more good news, in that the overall analysis leads to relevant strategies for handling various ill-conditioning issues that can arise in conjunction with Laplacian inverse source problems in these other areas. 

\begin{center} 
 \hfil\\
\textbf{\underline{Conceptual Preview}}
 \hfil\\
\end{center}

  For any true measure of success over a broad range of relevant problem ares, a surprisingly large body of diverse ideas and corresponding techniques will have to be considered here.  This led to inevitable difficulties in attempting to organize the underlying material.  In particular, since there are various issues that are hard to discuss in a straightforward way using a completely linear sequential exposition, it seemed appropriate to set the stage in several different ways.  Thus, many of the deeper issues subsequently encountered are raised in the remaining part of this ``Preamble Section'' and in the next ``Overview Section,'' prior to the more standard ``Introduction Section.''  Also difficult choices had to be made about what to leave out.  In particular, since geoexploration and physical geodesy are the primary focus here and the material was chosen with an eye towards generality, some significant special topic issues are not addressed at all---for example, that of handling conductors in the field region for electrostatic inverse source problems or that of adapting the formalism to accommodate the magnetic dipole form (versus the electrostatic or gravitational form that is more easily dealt with in DIDACKS theory).  In the end, the material that is contained in this article should be accessible to most of the readers from various other disciplines who might wish to try adapting it.  It goes without saying that, in the end, the reader will have to determine what concepts are relevant and appropriate for his or her problem area of interest.  In order to help orient general readers a brief survey of some of some of the main points made in the sequel will be given next.  (The order followed here does not necessarily follow that of the discussion in the main body of the paper.) 

 First, due to the fact that under the influence of gravity matter always attracts matter, the gravitational field energy is negative instead of positive.  This is obvious since energy is released when separated bits of matter come together to form, say, some planetary or stellar body.  (Conversely, the electrostatic energy density is positive and energy is required to assemble a charged body, since like charges repeal.)  For less massive bodies, such as the Earth, internal stresses and/or pressures restrain the resulting static configuration so that it does not implode entirely in upon itself, but the final configuration may well be a quasi-static configuration, where energy transfer of some sort or other must be considered.  (Obviously, tides, earthquakes and related terrestrial phenomena are manifestation of such energy transfer processes.)

 Next consider a concrete example of this gravitational body formation process.  If a massive (but not too massive, so that black hole formation is avoided) isotropic isolated non-rotating cloud of gas coalesces into a spherical ball under the action of mutual attraction of its parts, then, at some point in time, internal pressure will restrain the configuration from further collapse.  Even this simple example turns out to be fairly complicated, because time dependent thermal gradients and their effects on pressure must be taken into account.  Thus the core of this configuration becomes heated as it initially compresses and this, in turn, influences the pressure of the core itself; moreover, subsequent radiative cooling will have a strong influence on the final density profile and on how soon a stable (or nearly stable) configuration is obtained.  Detailed technical efforts to model planetary density configurations in terms of rotating gasses and incompressible fluids (as well as other forms of matter) have a long history and frame a pertinent part of astrophysics and geophysics \cite{Todhunter,Jard}.  For standard planetary bodies, these resulting axially symmetric configurations correspond to an ambient density profile that is  homogenous at each depth (i.e., tangentially isotropic) and these resulting nominal configurations can be taken as defining a minimum energy or ground state density configuration.  It is thus natural to remove this nominal reference field (which is labeled ``normal gravity'' by geophysicists \cite{HandM}) along with the included rotational effects and thus consider only density and gravitational field deviations from this equilibrium ``ground state.''   Clearly these density differences from a nominal (or normal) configuration result are as likely to be negative as positive.  The point being that these density differences indicate some sort of dislocation of matter that is associated with internal stresses and strains, so that they correspond to states of higher energy, which, in turn, means that the resulting external field differences also correspond to states of higher energy.  It is thus clear, then, that when a suitable reference field is subtracted, and one is willing to express density estimates as excursions from nominal conditions, that the proper strategy corresponds to minimizing the external field energy, subject, of course, to matching the available data (in, say, a point-wise sense).  As noted in \cite{DIDACKSII} this is precisely what DIDACKS point mass fits do:  They are the collocation fits that minimize the overall (weighted) field energy subject to the constraint that the potential be matched at specified data points. [The DIDACKS approach also simultaneously minimizes the (weighted) field energy of the difference between some specified truth field and the field to be estimated, which, in itself gives a strong added motivation for subtracting off a nominal reference field.]   

 As discussed in \cite{DIDACKSII}, it is worth noting that this method of residual fitting, as it is referred to in \cite{DIDACKSII}, is very useful within a general geophysical context.  In particular, as pointed out at the end of the main body of that article (\cite{DIDACKSII}), geophysical techniques generally always include subtracting off a standard ellipsoidal reference model from the total field and dealing with a quantity that is called gravity anomaly (or a related quantity called gravity disturbance) \cite{HandM}.  Often a further subtraction, which is called remove-and-restore (and that goes back to Fosberg, circa 1984), is carried out, which results in a localized field.  At several other places in \cite{DIDACKSII} the advantages of residual fitting techniques in DIDACKS based applications were also pointed out.  

  Second, a degree variance analysis shows that when a spherical harmonic reference field of degree and order nine or higher is subtracted off, then the spherical DIDACKS weighted energy norm relationships for the part of the field that is left give almost identical results (i.e., norms) to unweighted energy norm expressions for spherical exteriors; consequently, when a suitable reference field is subtracted off, the interpretational issues that arise for spherical exteriors associated with the difference between weighted energy expressions and direct energy expressions can be ignored.  It is also worth noting that this removal of a reference field has the effect of partitioning the density estimation problem into roughly two parts as well: (1) Core and deep mantle density estimation that is primarily tied to the chosen reference model chosen. (2) Geoexploration and other surface oriented geophysical density estimation areas that are primarily concerned with the remaining residual fields.  The concepts and techniques presented here are probably of most direct interest in geoexploration and other near surface related problem areas, but they clearly can also be adapted to the core and deep mantle regimes as well.

 Third, while there is clearly much more analysis that can, and should, be done along similar lines---especially with regards to the connections of core and mantle density distributions, along with all of the other geophysical and geodynamical aspects---in Section~\ref{S:inverse} certain commonly used regularization criteria are shown to correspond to assuming a direct proportionality between gravitational field energy and matter dislocations and an analysis is presented which shows that this connection has a reasonable physical interpretation.  Most of the other procedures that improve condition number correspond to assuming smoother density variations at the expense of choppy ones, which is one of the major overall themes of this paper.

  Finally, as an aside, although various types of point sources were considered in this and subsequent articles (including point dipoles and point quadrupoles), attention here is focused on theoretical and practical issues associated with mass density estimation from gravitational potential field information. 
Researchers in other disciplines that deal with inverse solutions to Laplace's (or Poison's) equation requiring dipole or other distributions as solutions can easily adapt the techniques presented here to their venue, so this is not nearly as restrictive as it may at first seem.

\begin{center} 
 \hfil\\
\textbf{\underline{Implementation Notes}}
 \hfil\\
\end{center}

  While, for most readers, it may be tempting to only ponder the various issues raised by this article, in order to have some true understanding of the issues raised, it is probably necessary to perform at least some numerical trial implementations.  A basic DIDACKS point mass fit is very easy to perform.  For example, in the half-space setting all that is necessary to do is to set up and solve (\ref{E:mkeqn}) using, say, the Householder triangulation routine in Lawson and Hanson \cite{LandH1}.  Since this article discusses and tries to justify various implementation points primarily from a conceptual point of view, it may not be clear how simple and easy to implement many of the suggestion are.  The goal of the remaining part of this section is thus to point out, by means of a few simple suggestions, how someone who is approaching these issues for the first time can gain some hands-on implementation experience in a relatively painless way.

 First, observe that since a DIDACKS point mass fit [i.e., $V_{N_p}$ specified by (\ref{E:pmpot})] satisfies the minimum collocation norm property, it is the potential with the smallest norm that matches the given (collocation point) data values of the specified potential $W$.   This means, for example, that if $V_{N_p}$ consists of a few shallow point masses only, then $V_{N_p} \rightarrow 0$ away from the specified collocation points, while if one adds more point masses and places them deeper the condition number will rapidly increase.  A standard way of improving the condition number is to add a regularization factor that is a quadratic function of the parameter set to the cost function.  For example, here the chosen modified cost function form might be
\begin{equation}\label{E:modPhi}
\Phi = \|V_{N_p} - W\|{\ls}_{{\text{E}}_1}^2 + \tau\sum\limits_{k=1}^{N_p} {m_k}^2\,,
\end{equation}
where $\tau$ is a small adjustable constant.
Observe that this modified cost function only tends to exacerbate the tendency of $V_{N_p}$ to ``regress to the mean'' since, for sizeable values of  $\tau$ it rapidly drives the values of $m_k$ to zero.  Notice, however, if prior to performing the fit some given reference model has been subtracted off from $W$, then the tendency of $V_{N_p}$ to ``regress to the mean'' implies that $V_{N_p}$ will regress to values that are relative to some preselected reference model, which is generally a desirable property.  Moreover, in the sequel it will be argued that under these circumstances one can conclude that this quadratic form is an appropriate expression for the potential energy that is associated with internal material dislocation. This subtraction of a reference model is called residual fitting in the sequel, just as it was in \cite{DIDACKS}.  It should be a simple matter for the reader to test all of this out for him or herself.

  Suppose that no reference model is available for use in residual fitting: What then?  Along simular lines that lead one to  conclude that quadratic expressions like the one on the right hand side (RHS) of (\ref{E:modPhi}) are associated with minimum energy states, on can argue that, in general, the smoother the density profile, the lower the energy state.  All of these considerations lead to a regularization factor of that is proportional to $\rightarrow \sum\limits_{i=1}^{N_p}\sum\limits_{j=1}^{N_p}  \omega_{i,\,j}({m_i - m_{j}})^2$, 
where the $\omega_{i,\,j}$ are positive constants that are nonzero only for source points that are not separated too much.
(Which clearly  makes the overall density profile more or less smooth.) Again the interested reader can test this out directly using simple examples.

  Finally, it is worth noting explicitly that the DIDACKS formalism does not have a built in way of handling error measurements, like geophysical collocation does \cite{Moritz}, so some sort of preprocessing is necessary when measurement errors are present.  As indicted in \cite{DIDACKSI}, a certain amount of caution is generally necessary---especially when using raw data.  To evaluate a candidate implementation, it is generally necessary to use some sort of realistic synthetic  data and perform a Monte Carlo analysis to get a handle on  the error behavior of the chosen implementation. (Even though standard covariance-based data preprocessing algorithms may give internal estimates of data statistics, they may not be completely reliable when, for example, downward continuation is present.)

 \vskip .35in
\noindent
 \begin{Large}{\textbf{(ii)\ \ Overview}}\end{Large} 

\vskip .18in
\noindent
  Given the general historical and physical import of Laplace's and Poisson's equation, methods of solution for either one in some particular realm are of general interest since the methods employed may touch on solution techniques in many other problem areas.  Thus while this article focuses on interior mass density reconstruction from given exterior gravity field information, many of the physical and mathematical strategies introduced here are quite general and can be either used directly (such as for gravity field modeling problems) or extended for use in other areas of applicability.  The basic philosophy used here in approaching density estimation problems is energy minimization because physical systems in stable equilibrium are clearly minimum energy states, which means that an energy minimization based approach can be used to by-pass well-known theoretical issues of ill-posedness that are generally linked to Laplacian (or Poisson Ian) equation density estimation problems. 

  This work builds directly on a mathematical framework presented in a previous article by the author that allows for the approximation of \riii\ harmonic fields in unbounded domains by (point) sources contained inside the (bounded) complimentary interior region \cite{DIDACKS}.  At the heart of this approach is the idea of a Dirichlet-integral dual-access collocation-kernel (DIDACK) that has the ubiquitous form of the inverse distance between some field variable point and a fixed source point.  Because the field point and source point are assumed to be in disjoint regions this kernel form is bounded; furthermore, this restriction to disjoint domains circumvents theorems disallowing reproducing kernels that have this general form and, in fact, these kernels can be employed to obtain closed form expressions for energy norm inner products.  Nevertheless, the resulting structure cannot be considered a reproducing kernel nor is it even a symmetric kernel since its two arguments do not share common domains.  There are, however, natural connections of the associated space (DIDACKS) techniques, reproducing kernel Hilbert space (RKHS) techniques and especially geophysical collocation, which is a specialized  \riii\ reproducing kernel technique \cite{Moritz}.   The RKHS and GC connections to DIDACKS theory, as well as the general mathematical backdrop and various precedents were addressed in this previous article that also dealt with the \riii\ half-space, in addition to the spherical exterior setting \cite{DIDACKS}.  It also included an overview of relevant aspects of physical geodesy and a brief outline of the author's, as well as others, experience with point mass fitting.  For the reader's convenience, the basic DIDACKS approach is reviewed in Section \ref{S:DIDACKS}, but since no attempt is made in Section \ref{S:DIDACKS} to motivate or re-derive the basic DIDACKS mathematical relationships, in what follows it is assumed that the reader is familiar with the overall plan of approach.  Connections of DIDACKS theory to various other mathematical approaches, such as the method of fundamental solutions, also exist and were addressed in a separate article  in this series dealing with DIDACKS \R{2}\ and \C\ theory \cite{DIDACKSI} and it is worth noting explicitly that many of the ideas developed here can be applied (or adapted) for use in these other mathematical settings.  

  Underpinning the mathematical and physical basis of the approach to Laplacian inverse source theory presented here are the DIDACKS closed form expressions for gravitational field energy, which yield a consistent source estimation procedure and interpretation when supplemented with four realizations that are central to this presentation:
\begin{enumerate} 
 \item
   The method of residual fits 
     \begin{itemize}
        \item  
          subtracts off a nominal density profile so the results are as likely to positive as negative,
        \item
          implies that any deviation from this nominal density profile are associated with an increase in field energy,
        \item
          geographically localizes the source estimation problem so that the procedure can be readily adapted to the geometry and data at hand.
      \end{itemize}
 \item
   When deviations from nominal conditions are under consideration, one can show that commonly used regularization procedures lead to a self-consistent physical interpretation and approach.
 \item
   Also conferring various implementation advantages, including greatly improved system condition numbers, is the structured point source technique (SPST), where groupings of point masses are used (with each grouping often being some selection of nearby point masses, generally taken to lie on a regular grid) and all of the masses in each of the groupings have predetermined relative mass values so that only an overall mass scale factor for each of the groupings is determined.  (Each of these groupings can also be regarded as defining a basis function.)  This has the effect of replacing a given point source basis function with a distribution that is more uniform and spread out, not only in terms of its density representation, but also in terms of the effective potential and forces produced.  As discussed below, one significant advantage of utilizing SPST basis functions is that a (family of) SPST basis function can be engineered to have  characteristics that meet preselected requirements, in say the frequency or spatial domain.  
\item
 Finally, since the DIDACKS energy minimization approach is based on a cost function, it can be seamlessly integrated with other cost function based approaches due to the inherent additivity of all cost function based approaches.  Moreover, alternative physical descriptions may easily be used to describe information content that was missing in the original cost function description: $\Phi_{Tot} = \Phi_A + {\tau}_B\Phi_B + \cdots$, for ${\tau}_B > 0$.  \{The point is that if the process described by $\Phi_A(\vec{\alpha})$ is physically consistent with that described by $\Phi_B(\vec{\alpha})$, for some (global) parameter set $\vec{\alpha}$, then the minimum of $\Phi_{Tot}(\vec{\alpha}) =$ minimum of $\Phi_{A}(\vec{\alpha}) =$ minimum of $\Phi_{B}(\vec{\alpha})$, while if the physical descriptions $A$ and $B$ are not consistent then $\tau > 0$ should be chosen in such a way as to reflect the relative reliability of $\Phi_A(\vec{\alpha})$ and $\Phi_B(\vec{\alpha})$ [in some instances the ratio of $\Phi_A(\vec{\alpha})$ and $\Phi_B(\vec{\alpha})$ (or its inverse) may be a direct expression of this relative reliability and suffices, as a general practical guide, for implementations]. 
 \end{enumerate}
The goal of this article is to articulate, clarify and amplify on these four points, while putting them within the overall context of a field energy minimization approach.  Given that the exact nature of these various points, how they relate to each other and how they relate to field energy minimization may be unclear at this stage (especially with regards to items 2, 3 and 4), a few additional side comments are in order here.

  It has already been noted that items 1 and 2 are connected.
 Next, observe that items 2 and 4 are related since the general way regularization  is added is by minimizing $\Phi(\vec{\alpha}) + \tau\,\bar{\Omega}$ instead of $\Phi(\vec{\alpha})$.  Here $\tau$ is an adjustable constant and $\bar{\Omega}$ is the so called ``regularization function.''  (Usually an over bar is not used in denoting the regularization function $\bar{\Omega}$, but an over bar is used here since the symbol ${\Omega}$ is reserved to represent the field region of interest).  Generally a regularization function (such as in Tikhinov regularization) is chosen solely for its condition number improving properties so $\tau$ is chosen to be as small as possible, consistent with this overall goal.  Here, however, the perspective is that ill-posedness is most likely a direct result of ignoring pertinent physical information about the underlying processes; consequently, the regularization process might be labeled ``constitutive regularization'' (versus Tikhinov regularization).  (It is perhaps worth noting, that for some years alternative information based approaches to inverse source theory have been suggested \cite{Tarantola,Tarantola2}).  That is, as subsequently argued, the point is that energy is generally associated directly with internal source dislocations and that, from a general constitutive perspective, there is a direct correspondence between often used regularization forms and reasonable expressions for this constitutive energy; conversely, simple assumptions and a straightforward analysis of the nature of this constitutive energy leads to natural forms of regularization functions.  It is also worth perhaps noting that a cost function based approach occasionally affords a easy means of collaboration.  In particular, with regards to item 4, although experts from respective fields A and B may have knowledge of their own specialty only, they may be able to form a collaborative effort where each submits his or her own separate cost function for use in the final total cost function, yielding a unique  composite optimal solution as a result.  This point is germane since the formalism presented in Section~5, when implemented along the lines of the last example given in this section, should allow for the tight integration of seismology data and gravitational data.  Finally, what is partitioned via 1 can also be added back by 4, so that there can be a subtle process of refinement of the total solution for certain relevant physical processes.

   Consideration of a few explicit examples may perhaps be necessary in order to clarify the main ideas behind item 3.  Thus, first consider a source density region partitioned into a set of non-overlapping cubes.  Although, in geoexploration, a collection of cubes (or parallelepipeds) such as this proves to be a very useful ansatz for gravity source estimation, there is an overall added level of implementation complexity due to the fact that the closed form potential (and gravity) expressions that result for each of the cubes [from integrating (\ref{E:IntPoisson}) below] are quite messy and this, in turn, clearly complicates any DIDACKS implementation (although such types of applications are not entirely unreasonable to consider---see the last part of Section~5, for example).  When the RHS of (\ref{E:IntPoisson}) is taken over a cube (say) the resulting potential function on the left-hand side (LHS) can be regarded as simply defining ``a cube basis function.''  On the other hand, one might consider a numerical approximation to this integral where the continuous distribution is replaced by a uniform grid of point masses, each of equal (but unknown) mass.  A better way to think of this is simply to regard this grid of uniform masses, not as a numerical approximation to a continuous distribution, but as a distinct type of distribution, which is to say a structured point source (SPST) basis function (as indicated above in item 3).  In this case, a SPST basis function can be regarded as an approximation to a cube basis function and it can be translated from place to place, just like a cube can; although, obviously in this case, each SPST basis function must be indexed by a separate label indicating its location.  Notice that this SPST basis function has only one undetermined parameter scale factor and that the total mass of the cube it approximates fixes this scale factor (or this overall scale factor can obviously be fixed by performing a DIDACKS fit---as discussed next).

 There are several additional points to be made here.  First, DIDACKS linear equation sets for the source parameters that result from using SPST basis functions are exact, easy to implement and easy to solve.  For example, when compared to a simple point mass fit, only the appropriate sums over the cubes have to be added to obtain the governing exact linear equations set.  Second, the matrix size of the linear system to be solved is determined by the number of regions of the system to be modeled (i.e., number of cubes) and not the size of the basis function internal grid, because there is only one unknown source parameter per source region, or SPST basis function. Consequently, a very fine (internal) grid can be taken, if desired, without increasing the size of the linear system to be solved.  Third, it is the dimensions of the cubes themselves (i.e., how close the source regions, or SPST basis functions, are together) and not the spacing of the internal grid that determines the condition number of the source parameter system.  Thus the use of SPST basis functions corresponds to using a sort of ``internal'' or ``structural'' regularization and, as such, the SPST approach is directly related to other energy based regularization techniques (c.f., item 2).  (The point here is that stable forms of solid or liquid matter have a certain inherent ``energy of constitution'' associated with either their molecular or crystal lattice structures and so long as this basic constitutive nature of matter is taken into account the actual amount of energy involved here does not matter since, aside from variations due to energy of deformation, the internal energies of material constitution are constant and hence ignorable so that only energy scales directly associated with deformations or dislocations need to be considered.)  For example, instead of the collapsing cloud of gas or dust considered above, one might consider (a more realistic?) model of an assemblage of preassembled uniform density rocks or other objects, each of which, since it is a preassembled uniform clump of a given type of matter, has an innate internal energy of constitution. Under gravitation collapse, it is clear that, generally, the final state of  such an assembly of matter will be in a minimum energy state, provided that there are insufficient pressures to cause excessive elastic deformations, and that under these circumstances the end configuration can generally be expected to be unique.  Fourth, as hinted at already by this type of gravitation collapse example, the resulting software implementation of a SPST approach can be made very flexible so that it can be adapted to various shapes, sizes and types of objects (and corresponding source regions).

   This brings up the second type of SPST basis function implementation example, which shows that there is generally a connection between items 3 and 4.  For the sake of concreteness, consider a case where there are three layers of unknown, but uniform density, and that each of these layers can be approximated by a single SPST basis function (each of which has an irregular boundary, in general). Further, assume that each of these slabs has an underlying common uniform point source grid, so that in matching the point sources to their corresponding basis functions it is only a matter of saying what grid point falls into what slab.  It is then only a matter of deterring the overall (scaling) densities for each of the three slabs.  Suppose that $\vec{\mu} = (\mu_1,\,\mu_2,\,\mu_3)$ represents the three source parameters (i.e., SPST basis scaling factors) of interest and that $\vec{\eta}$ are a finite set of parameters that determine the location of the boundary surface between the first slab and the second slab and that all of the other surfaces are known (and thus fixed) by some other means. [The $\vec{\eta}$ might be, for example, representative surface points that determine a Junkins interpolation, or they might be surface spline points, or they might be, say, a set GC surface determination parameters (with the idea being that if the statistics of the surface height are known then these surface parameters can be used in (more or less) the same way that GC is used in performing geoid height estimation).]  Then there are two sub points: first, one can simply perform a DIDACKS procedure to determine the total parameter set $(\vec{\alpha}) = (\vec{\mu}\,\,|\,\,\vec{\eta})^T$ by minimizing  $\Phi(\vec{\alpha})$ via standard nonlinear least-squares (NLLSQ) optimization means.  Here, for example if variations in the surface are due to mechanical stresses in the slab (associated, for example, with flexure of the slab itself) then a direct energy cost function can be associated with these surface parameters and, in accord with item 4 (as well as item 2), an energy regularization cost function can be added to $\Phi(\vec{\alpha})$, thereby both improving the estimate and its numerical tractability.  Second, in accord with item 4, additional data source types for determination of $(\vec{\alpha})$ can be entertained  besides gravity data, provided they can be written directly as a cost form process (here one obvious example for geoexploration problems might be to include seismology data).  


\section{Introduction}\label{S:intro}

 First, the overall mathematical setting can be described in terms of the Earth's gravitational field ${\vec{G}}_{\smallindexes{E}}(\vec{X})$ over the unbounded exterior domain, $\Omega \subset \mriii$, where the mass of the Earth is assumed to be contained inside the compliment of $\Omega$, denoted ${\Omega}'$.  This vector field is derivable from a scalar field or potential $W_{\smallindexes{E}}(\vec{X})$:  $ \vec{G}_{\smallindexes{E}}(\vec{X}) = \vec{\nabla} W_{\smallindexes{E}}$, where a positive sign on the RHS, rather than a negative one, is used to conform to the usual convention adopted in physical geodesy \cite{HandM} (other conventions were noted in \cite{DIDACKS}, but are of no immediate concern here).  Given this assumed default linkage of vector field and scalar field, only scalar potentials and their sources need be considered in the sequel.  The Earth's gravity field arises from some source density ${\rho}_{\smallindexes{E}}(\vec{X}')$ that is contained inside a bounded source region, $\Omega_{S}' \subset \Omega' \subset \mriii$. (Primed variables will generally be contained in the source region and unprimed ones in the exterior region, so a prime has been affixed to the source region symbol.)  Thus Poisson's equation, ${\nabla}^2\,W_{\smallindexes{E}} = -4\pi\,{\rho}_{\smallindexes{E}}$, holds for the whole of \riii\ and Laplace's equation holds for $\Omega$ since ${\rho}_{\smallindexes{E}} \eq 0$ there: ${\nabla}^2\,W_{\smallindexes{E}}(\vec{X}) = 0$ for $\vec{X} \in \Omega \subset \mriii$.  The potential field in $\Omega$ and its density in $\Omega_{S}'$ are then linked by:
\begin{equation}\label{E:IntPoisson}
W_{\smallindexes{E}}(\vec{X}) = \iiint\limits_{\Omega_{S}'} \frac{{\rho}_{\smallindexes{E}}(\vec{X}')}{|\vec{X} - \vec{X}'|}\,\, d^3 X'\,,
\end{equation}
which is the integral form of Poisson's equation.  [The question of particular units to be chosen is by-passed here, so a constant factor may need to be inserted on the RHS of (\ref{E:IntPoisson}).]

  Within this overall mathematical context there are two significant broad historical areas of research to contend with: (A) The determination of the Earth's global density profile, which along with the determination of the Earth's shape can be said to comprise the central issues of physical geodesy \cite{HandM}. (B) Problems associated with density determination for more localized regions arising form petroleum and mineral geoexploration efforts.  Here (A) addresses either deep mass distributions or shallower densities that do not vary abruptly, while (B) deals with near surface densities, and regions of abrupt change are often of special interest.  In some sense, geodynamics \cite{Geodynamics,AppliedGeodynamics} addresses issues that span both of these scales since it deals with phenomena such as plate tectonics and earthquakes, but since only configurations in static equilibrium (i.e., non-time dependent ones) will be explicitly considered, these problem arenas are not be addressed here in any detail.  Note that ${\rho}_{\smallindexes{E}}$ and $W_{\smallindexes{E}}$ can be conveniently partitioned into two parts corresponding to (A) and (B): ${\rho}_{\smallindexes{E}} = {\rho}_{\smallindexes{A}} + {\rho}_{\smallindexes{B}}$ and $W_{\smallindexes{E}} = W_{\smallindexes{A}} + W_{\smallindexes{B}}$ and that this partition simplifies a host of related interpretational issues. 

Here (A) goes back to the origins of potential theory itself and already had a long associated history by 1900 \cite{Todhunter}.   The realization that the problem of attempting to estimate interior mass density profiles from exterior gravitational fields is ill-posed goes back to Newton himself who showed that a uniform spherical mass shell and a point mass at the center of this shell produce the same exterior field provided they both have the same total mass.  Here it is assumed that the relevant aspects of this part (A) global field can be captured in terms of spherical harmonic expansions.  Recent progress in this area has been spurred by deployment of advanced satellite systems, such as the ongoing dual satellite Gravity Recovery and Climate Experiment (GRACE) (as well as the GROCE mission)  \cite{HandM}.  As previously noted, when spherical harmonic expansions of degree and order 9 (or higher) are subtracted off, then the fitting results obtained from the spherical weighted energy norm (i.e., the integral norm) match those of the spherical energy norm very closely.   Furthermore, if some care is exercised, then when a  spherical harmonic field of degree and order 120 (or higher) is used as a reference and removed then the half-space energy norm can be used.
 
\section{Synopsis of DIDACKS Approach}\label{S:DIDACKS}

  This section briefly outlines the basic mathematical formalism developed in \cite{DIDACKS}, where the main focus was point source Laplacian field reconstruction problems (i.e., gravity field modeling and estimation problems).

  First, consider the general DIDACKS plan of approach.  This approach is based on minimizing energy based norms of the difference between some point mass (or more general point source) model potential $v(\vec{X})$ and some given canonical reference (or truth) potential $w(\vec{X})$.  This can be restated directly in terms of  minimizing some cost function $\Phi' = \| v - w  \|^2$,
where $\|\,\cdot\,\|$ is the (possibly weighted) energy norm of interest.   For point mass basis functions the potential model becomes
\begin{equation}\label{E:pmpot}
 v(\vec{X}) = G\,\sum\limits_{k=1}^{N_k} \frac{m_k}{|\vec{X} - \vec{X}'_k|}\ .
\end{equation}
In (\ref{E:pmpot}) $G$ is the Newtonian gravitational constant $\approx 6.672\times 10^{-11} \text{m}^3 \text{s}^{-2} \text{Kg}^{-1}$ \cite{HandM}.  Here it useful to introduce scaled versions of the potential functions and to denote them by capitol letters so that $ V = v/G$ and $ W = w/G$ so that the factor of $G$ need not be considered in the sequel.  The relevant cost function thus becomes (where $\Phi \eq \Phi'/G$):
\begin{equation}\label{E:cost}
\Phi = \| V - W  \|{\ls}^2 = \|V\|{\ls}^2 - 2\,(V,\,W) + \|W\|{\ls}^2\,.
\end{equation}
 In the DIDACKS approach, since (weighted) energy norms for field energy expressions are used, the problem becomes to minimize  $\Phi = \|V - W\|^2_{E_j} := \iiint_{\Omega_j} \mu_j |\vec{\nabla} V - \vec{\nabla} W|^2\, d^3 X$, where $\mu_j = \mu(\vec{X})$ is the weight function (which may be set to one).  Besides this norm, (weighted) energy inner products will also be needed in the sequel: $(V,\,W)_{E_j} := \iiint_{\Omega_j} \mu_j \vec{\nabla} V\cdot\vec{\nabla} W\, d^3 X$. 

\SubSec{General Mathematical Considerations}

The notation conventions of \cite{DIDACKS} are followed here. 
 Cartesian coordinates are used in the sequel: $\vec{X} = (x, y, z)^T \in  \mriii$ and arrows are used to denote \riii\ vectors, while $n$ dimensional vectors (for $n > 3$) are denoted by lower case bold letters and their associated multidimensional matrices are denoted by upper case bold letters.    Further, $\text{R}_{\sps}$ will denote the radius of the sphere associated with $\Omega_{\sps}$ and the coordinate origin will be chosen to coincide with the center of this sphere so that $\Omega_{\sps} = \{\vec{X} \in  \mriii\mid |\vec{X}| > \text{R}_{\sps} \}$.  Likewise for the half-space case, the origin will be chosen to be in the plane $\partial \Omega_1$ and the positive z-axis will be taken normal to the plane so that $\Omega_1 = \{\vec{X} \in  \mriii \mid z > 0 \}$.  These two geometries will be denoted $\Omega_j$, where $j = {\sps}$ or $1$.  (Observe here that the general visual shape of the subscript matches the shape of $\partial\Omega_j$ itself.)  One other aspect of DIDACKS theory is worth noting, before addressing mathematical preliminaries.  

   Here general relationships that hold for both geometries of interest ($\Omega_{\sps}$ and $\Omega_1$) will be considered. 
For $\vec{x} \in \Omega_j$ ($j = 0$ or $1$) consider a vector field, $\vec{G}(\vec{X})$, derivable from a scalar field $W(\vec{X})$:  $ \vec{G}(\vec{X}) = \vec{\nabla} W$, where $\vec{X} \in $ \riii\ and all the sources are assumed to lie in some bounded ``source'' region, $\Omega'_{S_j} \subset \Omega'_j$.  We restrict ourselves from now on to potential functions that fall off at least as fast as $1/r$ as $r \rightarrow \infty$ in $\Omega_j$. 

\SubSec {\riii\ Half-space ($\Omega_1$)}

For concreteness, first consider the minimization process in \riii\ half-space, $\Omega_{1}$.  Here $z > 0$ characterizes the region of interest ($\Omega_1$).  It is clear that we wish to minimize a quantity with the following energy-like form (where the factor of $8\pi$ has been inserted since it occurs in the electrostatic field energy expression):
\begin{equation}\label{E:genlsq}
\|V - W\|{\ls}_{{\text{E}}_1}^2 \eq \frac1{8\pi}\iiint\limits_{\Omega_1} |\vec{\nabla} V - \vec{\nabla} W|^2\,\,d^3\vec{X}\,. 
\end{equation}
Here in general $\|V - W\|{\ls}_{{\text{E}}_1}^2 \eq \|V\|{\ls}_{{\text{E}}_1}^2 + \|W\|{\ls}_{{\text{E}}_1}^2 - 
 2(V,\,W){\ls}_{{\text{E}}_1}$, where, of course, the energy inner-product expression introduced after (\ref{E:cost}) is to be used for $(V,\,W){\ls}_{{\text{E}}_1}\,$.

   In particular if $V = V_{N_k}$ is a point mass model of interest (with $N_k$ masses) and $W$ is an appropriate given field, then $V_{N_k} = \sum_{k=1}^{N_k}m_k/|\vec{X} - {\vec{X}}'_k|$ where $\vec{X} \in \Omega_1$ and $\vec{X}'_k \in \Omega'_{S_1}$.  Further if ${\ell}_k \eq |\vec{X} - {\vec{X}}'_k|$, then
\begin{equation}\label{E:llsqeqn}
\|V_{N_k} - W\|{\ls}^2_{{\text{E}}_1} \eq \|W\|{\ls}^2_{{\text{E}}_1} -  2\sum_k^{N_k} m_k({\ell}^{-1}_k,\,W){\ls}_{{\text{E}}_1} + \sum_k^{N_k}\sum_{k'}^{N_k} m_k\,m_{k'}({\ell}^{-1}_k,{\ell}^{-1}_{k'}){\ls}^2_{{\text{E}}_1}\,.
\end{equation}
Observe that the first term on the RHS of (\ref{E:llsqeqn}) is a constant term.
Taking the partial of Equation~(\ref{E:llsqeqn}) with respect to $m_{k''}$ for $k'' = 1,\,2,\,3,\,\ldots,\,N_k$ and dividing by two yields a linear equation set that can be easily inverted for the $m_k$ values, provided the required inner products can be easily computed:
\begin{equation}\label{E:mkeqn}
 \sum\limits_{k'=1}^{N_k}T_{k, k'}\, m_{k'} = A_k,
\end{equation}
where $T_{k, k'} = ({\ell}_k^{-1},{\ell}_{k'}^{-1}){\ls}_{{\text{E}}_1}$ and $A_k = (W,{\ell}_k^{-1}){\ls}_{{\text{E}}_1}$\,.
 
   The DIDACKS formalism allows for the explicit closed-form evaluations of all the inner products occuring in (\ref{E:mkeqn}).  In particular, the energy inner product in this case is
\newcommand{\infBOTlim}{{\text{\begin{small}{$\!-\infty$}\end{small}}}}
\newcommand{\infTOPlim}{{\text{\begin{small}{$\infty$}\end{small}}}}
\newcommand{\smallW}{\text{\begin{small}{$W$}\end{small}}}
\newcommand{\smallg}{\text{\begin{small}{$g$}\end{small}}}
\begin{equation}\label{E:dirform}
({\ell}_k^{-1},\,W){\ls}_{{\text{E}}_1} \eq \frac1{8\pi}\iiint\limits_{\Omega_1} \vec{\nabla} {\ell}_k^{-1}\cdot\vec{\nabla} W\,\,d^3\vec{X}\ .
\end{equation}
and, as shown in \cite{DIDACKS}, 
\begin{equation}\label{E:e2rk}
({\ell}_k^{-1},\,W){\ls}_{{\text{E}}_1} = {W(x'_k,\,y'_k,\,-z'_k)}/4\,,
\end{equation}
which can be used to evaluate the inner product terms $T_{k, k'}$ and $A_k$ in (\ref{E:mkeqn}):
\begin{equation}\label{E:E1TKAK}
  T_{k, k'} = \frac14\,\frac1{\sqrt{(x'_k - x'_{k'})^2 + (y'_k - y'_{k'})^2 + (z'_k + z'_{k'})^2}}\,,\ \ \ A_k = \frac{W(x'_k,\,y'_k,\,|z'_k|)}4\ .
\end{equation}

\SubSec{\riii\ Spherical Exterior ($\Omega_{\sps}$)}

   Here $\Omega_{\sps} = \{\vec{X} \in  \mriii\mid |\vec{X}| > \text{R}_{\sps} \}$ describes the region of interest; however, matters are more complicated than they were for the half-space case.
  First consider two general admissible functions $f$ and $g$ (that is functions that are harmonic in $\Omega_{\sps}$ and which tail off to infinity at least as fast as $1/r$).   The energy inner product in this case is
\begin{equation}\label{E:E0norm}
 (f,\,g){\ls}_{\text{E}_{\spss}} \eq \,\frac1{8\pi}\negthinspace\iiint\limits_{\Omega_\sps} \vec{\nabla} f\cdot\vec{\nabla} g\,\,d^3\vec{X}\ . 
\end{equation}
The inner product for the ``integral norm'' \cite{DIDACKS} is also very useful here:
\begin{equation}\label{E:intnorm}
(f,\,g){\ls}_{\text{I}} \eq \,-\,\frac{R_{\sps}^2}{4\pi}\negthinspace\iint\limits_{\sigma} \Dr (f\,r\,g)\negmedspace\,\, d\,\sigma  = -\,\frac{R_{\sps}^2}{4\pi}\negthinspace\iint\limits_{\sigma} \big[ \Dr (f\,r\,g)\big]{\Big|}_{r=R_{\sps}}\negmedspace d\,\sigma
\ ,\ \text{ where } \Dr \eq \frac{\partial\ }{\partial r}\ 
\end{equation}
and the RHS of (\ref{E:intnorm}) follows from the evaluation convention given by
\begin{equation}\label{E:sigeqn}
 \iint\limits_{\sigma} f(r, \theta,\,\phi)\,d\,\sigma \eq \int\limits_{\theta=0}^{\ \ \theta=\pi}\negthickspace\negthickspace\!\!\int\limits_{\phi=0}^{\ \ \ \ \phi=2\pi} \negmedspace\!\!\!\left[f(r,\,\theta,\,\phi)\right]{\big|}_{r=R_{\sps}}\negmedspace\,\, \text{sin}(\theta)\,d\,\theta\,d\,\phi\
\end{equation}
for standard polar coordinates $r, \theta, \phi$.  Here, as in \cite{HandM}, $\sigma$ and $d\,\sigma$ have the standard meaning when associated with the integrand $f(\vec{X})$.  Likewise, let the surface inner product be defined as $$(v,\,w){\ls}_{\sigma} \eq (1/{4\pi})\,\iint_{\sigma} v(r, \theta,\,\phi)\,w(r, \theta,\,\phi)\,d\,\sigma\,.$$ 
With these definitions it is fairly easy to show \cite{DIDACKS} that
\begin{equation}\label{E:rel2}
 (f,\,g){\ls}_{\text{I}} = 4\,R_{\sps}(f,\,g){\ls}_{\text{E}_{\spss}} - R_{\sps}^2(f,\,g){\ls}_{\sigma}\ .
\end{equation}

  For DIDACKS applications over $\mathbb{R}^3$ spherical exteriors the integral norm is more important than the energy norm since closed form inner products can easily be computed from the following expression
\begin{equation}\label{E:IntRep}
(f,\,g){\ls}_{\text{I}} \, =\, {P_k }\,  W\left({\vec{P}_k}\right)\,, 
\end{equation}
where $P_k = |{\vec{P}_k}|$, with 
\begin{equation}\label{E:mkeqn3}
\vec{P}_k = \left(\frac {\text{R}_{\sps}^2}{|\vec{X}'_k|^2}\right) \vec{X}'_k
\end{equation}
for some point mass location $\vec{X}'_k$.  Here, the integral norm can be reinterpreted as a weighted energy expression \cite{DIDACKS}
\begin{equation}\label{E:inorm2d}
(f,\,g){\ls}_{\text{I}} =
\frac{R_{\sps}^2}{2\pi}\negthinspace\iiint\limits_{\Omega_{\sps}} r^{-1}\, \vec{\nabla}f\!\cdot\!\vec{\nabla}g\,\,d^3\vec{X} 
\end{equation}
so that $\mu_{\sps} = R_{\sps}/r$ ($\,r \eq |\vec{X}|$) is the associated weighting factor.

\section{Inverse Source Theory Prologue}\label{S:InverseSource}

    Petroleum and mineral and geoexploration are ongoing and historically significant research areas, where a considerable amount of time and effort has gone into exploring various alternative approaches and there is an extensive associated literature.  When a source distribution of interest produces a well delineated signal that can easily be separated from the background distribution, it is possible to simply compare the resulting potential pattern with some precomputed one.  Historically, this ``forward solution'' technique has been (and is) popular and it was probably the one first used \cite{OldRef}.  However, in the literature, when more sophisticated approaches are called for, the issue of the proper gravitational source estimation algorithm to use immediately becomes less clear.   While most of the approaches seem to work, all the currently used sophisticated approaches in this area entail a certain amount of arbitrariness or lack of physical motivation, which seems to be inherent in the foundations of all the approaches.

  Conceptually, in terms of the $V$ and $W$ above, one might frame the ideal goal as being to directly minimize $\Phi = \iiint_{\Omega_j^{\prime}} (\rho_V - \rho_W)^2\, d^3 X$, where $\rho_V$ and $\rho_W$ are the modeled and reference source terms.  However, when only $W$ and its derivatives are known in $\Omega_j$, there is no apparent way to effectively frame this minimization goal in a workable fashion.  This should be obvious from the fact that a continuous density like $\rho_W$ is generally nonunique, which, in turn, is clear from the fact that all spheres in a given region that have the same total mass and that share the same center produce the same external field---regardless of their radius.  As noted above, what is generally overlooked, however, is the fact if one takes into account the internal energy of the medium that is associated with stresses and other physical processes a local minimization of total energy will (generally) result in a physically unique situation (i.e., density estimate), because all physical problems in a static-stable equilibrium have a energy minimum underpinning.  Those situations that do not have a unique energy minimum are of special geophysical interest since they generally represent earthquakes, tides, core rotation or other geophysical situations, where dynamics (and the energy forms associated with it) must be considered.  All of this was discussed at some length in the first two sections of this article [Section (i) and Section (ii)].  The solution to the foundations of inverse source problems proposed there involves utilizing the DIDACKS approach (to account for external field energy differences) in conjunction with augmented energy-like information added as a regularizing factor.  Residual fitting also pays a central role in the physical interpretation.  Energy as a basis for studying  earthquakes has been proposed by others.  The material stress energy models in these studies is often very detailed and goes far beyond the scope of what can be included here, but integrating the DIDACKS approach with these considerations is clearly an avenue that warrants future effort since the contribution of external gravity field energy has been generally ignored in this arena.  In the current paper a constitutive regularization approach is taken and the goals are much more technically modest.  The goal is simply to physically justify easily implemented internal energy minimization approaches, where flexibility and ease of implementation are maintained as a primary goals.

  It is worth explicitly noting that the general role of energy minimization obviously has not gone unnoticed historically.  In particular,  Kellogg \cite{Kellogg} explicitly points out the role or field energy minimization in electrostatics via Dirichlet's integral \cite[p. 279]{Kellogg}, but there are clearly many other historical connections that can be pointed out in this context.  It is also worth noting that when the foundations of geophysical collocation were debated by Krarup and Moritz, Moritz put forth a statistical interpretation (which is commonly called least squares collocation) that eventually won out, but that Krarup put fort the idea of a weighted energy minimization approach based, effectively, on the RHS of (\ref{E:inorm2d}).  As discussed in the last section of \cite{DIDACKS}, since the goal was to give an interpretation to GC his idea was to use the Krarup kernel (rather than $1/{\ell}_k$ used by DIDACKS theory) and to argue for the physical importance of energy minimization.  Part of this debate can be glimpsed from some side comments in early geophysical colloquium proceedings.  As pointed out in \cite{DIDACKS}, DIDACKS theory turns things around conceptually and abandons the pretext of a symmetric reproducing kernel (SRK), which largely severs the direct connections to geophysical collocation, while it keeps energy minimization and the fundamental of the form $|\vec{X} - \vec{X}'_k|^{-1}$ in tact.  In the end, as argued in this paper, this also has the effect of keeping connections open between density estimation and energy minimization.

\section{DIDACKS Implementation Discussion}\label{S:phys}

   As pointed out in Section~\ref{S:DIDACKS}, there are two field regions of interest, the exterior of a sphere and positive half-space, denoted by $\Omega_j$ (for $j = \sps$ and $1$), respectively.  The corresponding energy norm for these two regions can then be simultaneously referred to as $\|\cdot\|{\ls}_{{\text{E}}_j}$.  The corresponding DIDACKS norms for these two regions can likewise be referred to as $\|\cdot\|{\ls}_{{\text{D}}_j}$ where
\begin{equation}\notag
\|\cdot\|{\ls}_{{\text{D}}_1} \eq \|\,\cdot\,\|{\ls}_{{\text{E}}_1}\ \ \text{and}\ \ \|\,\cdot\,\|{\ls}_{{\text{D}}_{\sps}} \eq k\|\,\cdot\,\|{\ls}_{{\text{I}}}\ .
\end{equation}
Here $k$ is a constant which can be chosen to preserve connections to units of energy [i.e., $k = 1/(4R_{\sps})$]; however, since the resulting values of source parameters for a DIDACKS fit do not depend on the specific value of this constant, it is generally more convenient to simply set $k = 1$.  For $\Omega_{1}$, $\|\cdot\|{\ls}_{{\text{D}}_{1}}$ and $\|\cdot\|{\ls}_{{\text{E}}_{1}}$ are the same so the question of which to use does not arise; however,
for $\Omega_{\sps}$ it would appear that for inverse source problems there is some question as to whether $\|\cdot\|{\ls}_{{\text{E}}_{\sps}}$ or $\|\cdot\|{\ls}_{{\text{D}}_{\sps}}$ should be employed as the major tool of analysis; where, of course, $\|\cdot\|{\ls}_{{\text{E}}_{\sps}}$ has a direct bearing on assorted energy based arguments, but $\|\cdot\|{\ls}_{{\text{D}}_{\sps}}$ is more mathematically amenable.  Nevertheless, as previously pointed out, this dichotomy goes away for geophysical problems if a suitable low degree and order spherical harmonic reference is subtracted off (and then restored at the appropriate time), because the resulting  residual field has no low degree and order content and the two norms, in this case, are nearly proportional.  That this is so can be seen from a full spectral analysis.  That this is so can also be seen by taking stock of (\ref{E:inorm2d}), where it was already observed that $\mu_{\sps} = R_{\sps}/r$ is the effective weighting factor for the weighted energy norm in this case.  The point here being that after a reference field is subtracted off, the remaining residual field attenuates very rapidly as $r$ increases so that only values of the field close to $R_{\sps}$ make significant contributions to the norm and in this ``near field'' region $\mu_{\sps}$ is approximately constant.  

 There are two direct consequences of the DIDACKS (weighted) energy minimization approach:
\begin{itemize}
\item[(A)]  Since $\|V - W\|{\ls}_{{\text{D}}_{j}}^2$ is minimized, the resulting fit will be the one which minimizes the (weighted) energy difference of the error field [which by definition has the potential form ($V - W$)]. 
\item[(B)]  Since the GC property is satisfied the resulting field will be the one which also minimizes $\|V\|{\ls}_{{\text{D}}_{j}}^2$ subject to the constraint that the sample field data points be matched (which is the replication property, so that, for example, for point mass fits $V(\vec{P}_k) = W(\vec{P}_k)$ at all the specified data points $\{\vec{P}_k\}_{k=1}^{N_k}\,$). 
\end{itemize}
 While the interpretation of all this is all relatively straightforward, there are various issues that warrant consideration and further clarification.  The first issue to be reconsidered is the sign of the gravitational energy itself.

\SubSec {Negative Gravitational Field Energy and Residual Fitting}

   Amplifying slightly on the discussions in previous sections, the electrostatic case will be compared with the gravitation case.  Both electrostatics and gravitation are inverse square law forces and in both cases, for source free regions, the associated scalar potentials obey Laplace's equation.  In the electrostatic case, the forces between two charged bodies are proportional to the product of their charges, while in the gravitational case the forces are proportional to the product of their two masses.   Aside from the fact that masses are always positive and charges are not, which causes some minor interpretational issues here, there is one fact that cannot be ignored:  like charges respell and like masses attract, so gravitational forces are always attractive.  (The sign differences associated with the choice of the gradient of the potential may complicate the identification of gravitational potential and potential energy, but this is not a issue that needs to be addressed here.) The fact that gravity is always attractive means that the gravitational field energy is inherently negative, unlike the electrostatic case.  To see this consider what happens when a set of gravitational or electrostatic sources are assembled from an initial configuration that is well separated (i.e., out at infinity), which is, in general, how one computes the field energy.  In the electrostatic case, when a collection of like sources are assembled it is clear that positive work must be done to overcome the mutual repulsion of the charges.  From well-known arguments found in standard physics texts the resulting electrostatic field energy has the form 
$ = \text{constant}\times\tfrac1{8\pi}\iint \vec{E}\cdot\vec{E}\,d^3 \vec{X} > 0$.  However in the gravitational case work is released during the assembly process (which, as noted in Section~(i), is invariably gravitational collapse process) so that the work of assembly is thus $-\tfrac1{8\pi}\iint \vec{G}\cdot\vec{G}\,d^3 \vec{X} < 0$.

  Clearly the fact that the gravitational field energy is negative and thus related to the negative of $\|W\|{\ls}_{{\text{E}}_{j}}^2$ is of prime interest here, since this difference in sign, at a minimum, is somewhat unsettling from the requirements of a consistent physical interpretation.  First, however, notice that,
 as discussed in \cite{DIDACKS}, conditions (A) and (B) above, by themselves, provide sufficient direct motivation for handling modeling and estimation problems, since these conditions imply that the gravity error field is a minimum [by condition (A)] and that the fit is the most conservative one consistent with the given (point) data set [by condition (B)].  With regards to both conditions (A) and (B), notice that since the absolute value of the field energy is minimized, smaller (and therefore more conservative) overall excursions are emphasized over larger excursions.  Also observe that (A) implies that it is not the absolute value of the field energy itself that is minimized, but the positive absolute value of the energy of the error field itself and this is clearly desirable.  In general, as previously noted, residual fitting is a part of the DIDACKS approach, which means that $W$ usually represents not the raw gravity field itself, but a residual field where some suitable well defined base reference function has been subtracted off.  The resulting residual field can then be assumed to be zero-mean in the sense that when a point mass fit is performed on it the resulting point mass values occur in roughly equal positive and negative proportions, which is to say that $\sum_{k=1}^{N_k}m_k \approx 0$.  As pointed out in Sections~(i) and (ii), and backed up by analysis in Section~\ref{S:inverse}, this means that positive field energy is associated with (nonzero) source excursions, which provides a direct explanation of the negative gravitational field energy mystery noted above.  This also clearly provides a direct explanation for the possibility of negative mass values arising in conjunction with point mass fits.   Condition~(B) thus must really be considered as holding for a residual field, where all the field excursions are to be considered as excursions from zero, so that the energy of the difference fields can always be considered positive.  This means that the absolute value of the residual field energy is minimized subject replication constraints, which is clearly desirable. 

\SubSec {The Point Source Support Problem (PSSP)}

Clearly for standard point mass fits used in gravity modeling or estimation the DIDACKS approach does not generally require numerical discretization or numerical integration since the underlying linear equations are in a closed form; however, in this case ease of software implementation does not necessarily translate into uniformly care-free applicability.  For a straightforward point mass fit, depth and spacing issues must be handled more-or-less correctly in order to obtain acceptable results.   One common mistake made in utilizing point masses for modeling purposes is to not place them deep enough.   Consider a point mass model based on linear equation~(\ref{E:mkeqn}).  From the form of $\widetilde{T}_{k, k'}$ specified by (\ref{E:E1TKAK}) [where the overset tilde indicates the use of normalized basis functions] it is clear that as point masses are moved closer to the surface or further apart horizontally they clearly become less correlated (that is, $\widetilde{T}_{k, k'}$ becomes smaller).  In the limit that the point masses are all near the surface they match the prescribed potential values at the specified locations, but the given potential model itself falls off to zero very quickly at locations away from those specified data points.  In fact, under these near-surface circumstances a single point mass fit to a single data point behaves very much like a Dirac delta function.  This is clearly an undesirable situation and to overcome it one must place the point masses at a fairly sizeable depth.  Alternatively, as the point masses are moved closer together or placed deeper they quickly become overly correlated ($\widetilde{T}_{k, k'} \rightarrow 1$), which, in itself, can lead to wild and unexpected variations in the resulting field model at points away from the prescribed reference field data points.  These results clearly hold for the region $\Omega_{\sps}$ as well and they hold for other point source types, such as dipoles and quadrupoles, as well.  Often a fairly fine line between these two just outlined opposing and unwanted behaviors must be negotiated.   A good balance of spacing and depth must be struck and when potential data locations of various heights and spacing is involved, it can become a very difficult (or even nearly impossible) problem to overcome.  This problem is thus labeled the Point Source Support Problem (PSSP), because it has to do with support issues associated with the underlying point source basis functions themselves.  Clearly these problems are dependent on the choice of basis function, which is one reason that SPST basis functions were introduced in Section (ii).  As previously discussed, various other techniques can also be utilized to overcome these issues and much of the rest of this article will explain relevant aspects of them within one context or another.  The main means for overcoming the PSSP discussed in this section are residual fitting and spectral bandwidth decomposition.  Various types of regularization and various distributed types of sources will be considered in the context of inverse source determination theory.  Condition number considerations are closely tied to the PSSP and are also discussed below.

\SubSec {Residual Fitting and Spectral bandwidth Techniques}
 
 Next consider residual fitting as applied to gravity estimation, gravity modeling or source estimation, where it should be normally considered as an integral part of the DIDACKS approach in these areas.  There are three primary reasons that the residual fitting technique is so effective for point source problems.  First a certain number of degrees of freedom are always tied up in reproducing the general trends of the underlying reference model and when these reference trends are no longer present these additional degrees of freedom are freed up and can be used for modeling finer detail.   Since simultaneously fitting a rapidly changing gravity field (which tends to require shallow point sources) and a field with long term trend properties (which tends to require deep source placement) is often difficult at best, residual fitting can be used to eliminate much (or most) of the long term attributes to be fit so that the regional part of the fit becomes, not only much more accurate, but easier to effectively implement. (Thus helping to overcome PSSP issues.)  The second reason residual fitting is effective is associated with the collocation replication property, which DIDACKS fits obey.  As previously noted, for techniques satisfying the collocation property, the fit usually digresses to zero away from the specified (field) data points; however, when a reference is subtracted off, this natural digression will be to the underlying reference model itself so that there is a natural attenuation built in.  (This, in itself, clearly also helps alleviate point source support problems.)  The third reason will be addressed next by itself and has to do with consistency of physical interpretation and is tied to more general field energy considerations, as previously discussed.  

  After a residual fit has been performed to model the field to a certain physical scale, the entire process can be repeated and when such a series of residual fits is performed there is a synergistic effect.  First since the (weighted) energy norms tend to fit the longest wave lengths first, the first fits performed (naturally, with sources chosen to be at a greater depth) will account for that part of the field that tails off more slowly with altitude.   In turn, when this part of the field is treated as a reference and removed only the shorter wave length and more regionalized part of the field remains to be fit.  The whole process can then be repeated as needed.  In conjunction which this repeated residual fitting process note that it is important to remove the longer wave lengths present at each of the successive stages, or much of the error at each stage will be folded into the parts of the field that are be to fit subsequently.  In this connection, it is worth explicitly noting that for the integral norm a degree variance analysis (or harmonic Fourier series analysis for the half-space energy norm) shows explicitly that a strong premium is placed on correctly matching any longer wavelengths that happen to be present.  There is one negative aspect of residual fitting techniques.  Since residual fitting techniques work primarily due to preconditioning of the `signal' (so that it can accurately be fit by point sources), in general a good resulting point source fit will entail sources that are deeper than one might normally expect.  This, in turn, leads to associated condition numbers that are large.  If signal errors are present or if source estimation is the main goal, rather than modeling, clearly there may reason for concern.

  In many cases a spectral bandwidth approach can be used in place of residual fitting.  With regards to the global part of the field a spectral bandwidth approach simply entails dividing  up a  spherical harmonic expansion of $W$ into various degree ranges so that each resulting spectral band has well defined physical attributes.  Since such spherical harmonic reference fields are both accurate and readily available, it is assumed, that at the very minimum, that some low degree and order reference field will be subtracted off and used as a base reference for either a residual fit or spectral band approach.  For local or regional fits where data is specified at some survey altitude, one might use digital filtering techniques or a fast Fourier transform in order to obtain various spectral or frequency bands.  With sufficient ingenuity, interested readers should be able to figure out any required further details for such implementations so they are not discussed here.  There are, however, two further points that are worth commenting on in this subsection.
 
Considering the spherical case, it is natural to assume that the surface of the earth is to be taken as coincident with $\Omega_j$; however, when the data is specified at a fixed altitude (\ref{E:mkeqn3}) fixes the point source depth at corresponding depths that may be totally inappropriate for the associated frequency content.  Clearly one solution is to simply upward (or downward) continue the original data by using GC to an altitude that will produce an acceptable point source depth for the required spacing. In practice an alternative technique that generally works quit well is more appropriate.  For concreteness, consider a DIDACKS point source fitting problem based on the spherical exterior geometry using the spectral-bandwidth approach.   Further, suppose that data is given on the surface of the Earth, which is specified by $|\vec{X}| = R_E$.  (In practice, for a regional fit it is natural to take the origin to be directly under the center of the region along the ellipsoidal normal direction and at a distance that best captures the ellipsoidal curvature effects over the total region of interest.)   Next determine what the appropriate spacing and depth should be for an ideal fit.  The approach is then simply to consider $R_{\sps}$ to be a variable ($ < R_E$) that is to be set to a value that will insure that (\ref{E:mkeqn3}) produces this desired depth---for data sampled at the correct spacing.  This technique entails no loss of consistency since these deeper point sources are associated with a field region that may, in fact, be naturally taken to have a much smaller $R_{\sps}$.  In particular, there is no reason to ignore the (weighted) field energy between $R_{\sps}$ and $R_E$ produced by these deeper sources by arbitrarily taking $R_{\sps} = R_E$.  (In any case, for deep sources the source exterior region clearly has a boundary that is somewhat below $R_E$ and there is no real reason for thinking that $R_{\sps} = R_E$ is the correct boundary for weighted exterior field energy minimization over this part of the field.)  This same technique can also easily adapted to the geometry specified by $\Omega_1$.   Here it is a simple matter to move the plane $\partial \Omega_1$ deeper, which forces the associated point sources themselves deeper. 

  Finally, in the present context, it is interesting to note that a simular boundary adjustment technique to that just described can be used to produce a norm that minimizes energy  over a region bounded by two planes (or weighted energy over a region between two concentric spheres).  Although the technique is general it is easiest to describe it in terms of a single point source for the geometry $\Omega_{\sps}$.  Thus, suppose that $\Omega_{\sps} = \{\vec{X} \in  \mriii\mid |\vec{X}| > \text{R}_{\sps} \}$ and $\Omega^{\star}_{\sps} = \{\vec{X} \in  \mriii\mid |\vec{X}| > \text{R}^{\star}_{\sps} \}$ describe two spherical DIDACKS regions with the same origin and $R_{\sps} > R^{\star}_{\sps}$.  Let $\Omega_{\circledcirc} \eq \Omega_{\sps} - \Omega^{\star}_{\sps}$ be the region of interest, then from (\ref{E:IntRep}) there results:
\begin{equation}\label{E:DelEqn1}
 \text{D}[w,\,\ell_k^{-1},\,\mu_{\sps},\,\Omega_{\circledcirc}] = |\vec{P}_k|\,\,w(\vec{P}_k) - 
|\vec{P}_k^{\star}|\,\,w(\vec{P}_k^{\star})
\end{equation}
with $\vec{P}_k$ from (\ref{E:mkeqn3}) and 
\begin{equation}\label{E:DelEqn2}
\vec{P}_k^{\star} = \left(\frac {(\text{R}^{\star}_{\sps})^2}{|\vec{X}'_k|^2}\right) \vec{X}'_k.
\end{equation}
Clearly a like expression follows for the region bounded by two parallel planes.

\SubSec{Condition Number Considerations}

  Commonly available singular value decomposition (SVD) or Householder triangulation routines are appropriate for solving the DIDACKS point source determination linear equation sets.  (Generally, the amount of processor execution time is so minimal that it is not a real consideration and, thus, except for rare circumstances, the universal reliability of the solution from a SVD or Householder triangulation algorithm is of much greater importance than execution time.)  The data replication property allows one to verify that the implementation has been performed correctly.  The eigenvalues, which should be all positive, can be readily obtained from SVD routines.  (Here it is worth noting  that for those few cases where one might require $m_k > 0$ linear inequality constraint software can be employed, but note that one should generally validate the output of this software \cite{LDP}.)  The condition number ($C_\#$) is taken here to be the ratio of the largest to the smallest eigenvalue of the $\mathbf{T}$ matrix in this linear equation set.  

  As discussed above, for modeling problems, when the masses are too shallow the $C_\#$ will be too small.  When the masses are too deep the $C_\#$ will be too large.  For source determination problems matters are somewhat different and the $C_\#$ should be somewhat smaller.   The point is that a large condition number is generally associated with large variations (and thus uncertainty) in the estimated masses and this is obviously associated with uncertainty in the prediction of the mass density itself; moreover, a large $C_\#$ indicates that any measurement errors will tend to be magnified by a like amount.  Special techniques that lower the condition number, while simultaneously overcoming the point mass support problem, have been emphasized elsewhere in this article (obviously, many of these techniques are also appropriate for use in gravity estimation and gravity modeling problems as well).  A guiding principle is that the less certain one is about the fitting results the lower the $C_\#$ should be; moreover, for modeling problems in general one does not care about the mass values themselves---only the results.  This means that unphysical mass values are perfectly acceptable if they produce a good fit.  In this context, it is perhaps worth noting that for low degree and order global point mass NLLSQ fits good results are associated with $C_\# \gtrapprox 10^{10}$, but for point mass modeling with various fixed locations, one would generally expect somewhat smaller condition numbers than this threshold (results for grids over interior regions were noted in \cite{DIDACKSII}, but the same gridding techniques can obviously be profitably used for exterior regions as well).

  Before proceeding with the discussion of other fitting techniques in this next section, a word of caution is in order.  Some sort of experience with point mass fitting is probably required before attacking real world estimation or inverse source determination problems.  Thus, it is suggested that the reader interested in these areas gain as much experience through synthetic modeling as possible by working with various field models $W$, which are chosen to have properties that are as realistic as possible.
For estimation problems, such modeling allows one to check the produced field values by the intended spacing of the point set at various locations away from the field sampling points.  Likewise, for source estimation problems one can test the predicted source values against the ones used to produce $W$.

   Finally, it is also worth noting that the PSSP can be overcome generally by inputting a tolerance to the linear inversion software so that unwanted small eigenvalues are ignored (this is generally a very strong form of regularization), thus allowing a very tight point source grid spacing while preventing large source values.  Here is also worth noting that when normalized basis functions are used, the largest eigenvalue for the system (\ref{E:mkeqn}), or its spherical analog, is obviously bounded from above by $N_k$. 
 
\section{Inverse Source Determination Techniques}\label{S:inverse}

   For source determination problems $C_{\#}$ concerns must be addressed and there are two primary means of doing this: regularization techniques and basis function modification techniques. As just discussed in Section~\ref{S:phys}, for optimal fits, residual fitting techniques generally overcome the PSSP, but generally at the expense of large $C_{\#}$'s; however, residual fitting is still an important source determination technique, since it works synergistically with basis function modification techniques and regularization.  The primary types of alternative basis functions to point sources that will be considered are structured point sources (which consist of an aggregate of point sources).  Basis functions that yield continuous source estimates are also considered.  In particular, since continuous sources are inherently nonunique, the primary tool considered in this regard is parameterized continuous source estimation.  Other parameterized continuous source techniques have long been used in geophysical inverse source theory \cite{Parker}

\SubSec {Regularization Techniques}

 Before proceeding to the analysis of the physical basis of specific regularization approaches, it is useful to recap part of the analysis given in previous sections from a slightly different perspective. 

  As noted in Section~\ref{S:InverseSource}, for density estimation problems one would ideally like to minimize an expression like $\Phi = \iiint_{\Omega_{S_j}^{\prime}} (\rho{\ls}_V - \rho{\ls}_W)^2\, d^3 X$ where $\rho{\ls}_V$ and $\rho{\ls}_W$ are the modeled and reference source terms and $\Omega_{S_j}^{\prime}$ is the source region associated with the two field regions of interest ($\Omega_j$); however, there appears to be no way to mathematically frame this minimization goal when only $W$ and its derivatives are known in $\Omega_j$.  It is easy enough to see that this must be the case from purely information content alone since $W$ is harmonic in $\Omega_j$ and thus is specified by its values on $\partial\Omega_j$, whereas $\rho{\ls}_W$ has many more degrees of freedom and is not determined by its values on a surface.  As noted in \cite{DIDACKS}, for a chosen closed surface in $\Omega_j$, DIDACKS theory links specification of potential values to source values specified on a corresponding closed surface in the source region.  From this observation and the fact that equal mass concentric shells produce the same external field, one can conclude that much of the ill-posedness of the density problem are associated with source depth issues.  Two aspects, however, are clear: (1) Regularization, in any reasonable form, should help to stabilize the source estimates and thus generally provide more reliable estimates.  (2) Gravitational inverse source problems minimize energy in some sense or another since all physical systems in static-stable equilibrium have energy minimum underpinnings.   The general thrust here is thus to try to link these two aspects in the DIDACKS approach to source determination.  As previously noted, residual fitting is linked to the interplay of these two aspects.

   In order to ascertain some of the underlying issues involved, consider a straightforward application of the point mass fitting theory presented in previous sections (without regularization).  In this case the approach is based on the minimization of $\|V - W\|{\ls}_{{\text{E}}_j}^2 \approx \|V - W\|{\ls}_{{\text{D}}_j}^2$ for field information specified in the region $\Omega_j$ (for $j = \sps$ and $1$).  (In what follows the norm expressions will be written in terms of energy norms.)  As before it is assumed that an appropriate reference function has been removed from the specified function $W$ prior to fitting, which entails the removal of a reference density from $\rho{\ls}_W$ as well (but this removal may only be implicit).  Residual fitting helps here since residual fields have reduced low frequency content, which allows for shallower point masses placements of the remaining sources.  The point being that the ill-conditioning arising from the source depth ill-posedness mentioned above can be overcome by introducing a natural source depth stratification.  This helps to control one cause of innate ill-conditioning, but there is another one that is associated with how close the sources are together.  Thus, in general, the condition number will still be much too high as the grid spacing becomes small.  As previously noted, a large condition number is unacceptable in this case for three reasons:  (1) The mass values will tend to vary wildly from one point mass location to another, so a satisfactory limit is hard to obtain.  (2) A large condition number indicates a lack of knowledge in the inverse source determination process itself, so the predicted results will be questionable.  (3) Any data errors or extraneous high frequency content present will be greatly exaggerated in the source estimates.  These issues are clearly related to the associated Point Source Support Problem (PSSP).  Regularization can be used to largely solve these conditioning problems in a natural way. 

   To motivate what follows, consider a preliminary argument indicating a connection of mass dislocation and energy.  Toward that end, consider the rather specialized situation where a very detailed reference model exists that fully represents the part of the mass density that is locally homogenous so that all that is left to predict are local density irregularities.  Suppose, further, that this reference model has been subtracted from $W$ and that a small enough uniform grid spacing is used so that one can directly associate a given point mass value with local density irregularities, which can be reinterpreted as a small block of matter.  Two different physical scales will be considered, where the finer one is associated with this uniform grid of point mass locations.  (The point here is that generally one should distinguish between the fixed framework of point masses that are used for estimation purposes, which are generally assumed to be at fixed locations and the mass distributions that they are assumed to model, which may well shift.)  For location $k$, let $m_k$ be the point mass value in question and let $M_k$ be the total mass of this block associated with the given potential $W$.  If there are no local stresses in the medium, then the larger physical block of matter that $M_k$ is part of is in its normal configuration and the subtracted reference field accounts for all of the local density $\rho_k$ so that $m_k = 0$.  Alternatively, consider what happens when the larger block of matter that $m_k$ is part of is subjected to compression along one direction, say the vertical direction.  Let a capitol letter $K$ be associated with this larger physical block of matter that $M_k$ is part of so that its mass is represented by $M_K$.  Then, let $L_K$, $W_K$ and $D_K$ denote the length, width and depth of this larger block.  Further assume that the distortion of this larger block ($\delta D_K$) is small and that the block responds in an elastic (i.e., linear) manner to this force by a change in $D_K$ only.  Since the block mass is conserved, $M_K$ is constant and thus $\delta \rho_K = {M_K}{/(L_K\,W_K\,[D_K - \delta D_K])} - {M_K}/{(L_K\,W_K\,D_K)} \approx \delta D_K\,\rho_K$, so that $m_k \propto \delta \rho_K \propto \delta D_K$.  Finally since the potential energy associated with elastic forces is proportional to $(\delta D_k)^2$ it is clear that the energy of this internal dislocation is proportional to $m_k^2$ and thus the total energy for all the dislocations caused by all the various stresses can be written as $\sum_{k=1}^{N_k} \alpha_k\,m_k^2$, where the $\alpha_k$'s are constants of proportionality.  If all of the blocks can be treated consistently, then this energy can be written as $\alpha\,\sum_{k=1}^{N_k}m_k^2$.  Adding this internal configuration energy to the external field energy form yields
$$\Phi = \|V - W\|{\ls}_{{\text{E}}_j}^2 + \alpha\,\bar{\Omega}$$
 as a more accurate replacement for $\Phi = \|V - W\|{\ls}_{{\text{E}}_j}^2$, where
$$\bar{\Omega} = \bar{\Omega}_1 \eq \sum_{k=1}^{N_k}m_k^2\ .$$
This is a standard quadratic regularization form that is invariably introduced solely on the grounds that it reduces the condition number.   In particular, notice that using $\bar{\Omega}_1$ effectively adds a diagonal term to $T_{k,\,k'}$ and this clearly reduces the condition number of the linear equation set (which is especially obvious when normalized basis functions are used).

 The general philosophy underpinning the use of $\bar{\Omega}$ here is easily stated.  When minimization of $\|V - W\|{\ls}_{{\text{E}}_j}^2$ fails to specify a unique density estimate, the addition of $\bar{\Omega}$ will select those densities that have the lowest internal energy configuration, all other things being equal.  (It is generally accepted that the action of a regularization form like $\bar{\Omega}_1$ produces a unique fit.)  Finally before heading on it is worth discussing the implementation of normalized basis functions here.  The action of a regularizing function is generally ignored when the normalization conditions are implemented so that it is fixed regardless of regularization.  The regularization functions involving mass (like $\bar{\Omega}_1$), however, should be defined in terms of $\tilde{m}_k$ rather than $m_k$, but this complication is not considered in the text.  (Further it worth noting that it will be assumed in the sequel that normalized basis functions are used; however, for convenience this normalization process is generally carried out without considering $\bar{\Omega}$, and then the effects of this term are added in just prior to computation of the linear equation set solution.)

 While the reduction of condition number associated with the use of $\bar{\Omega}_1$ is desirable, there are three additional complications to consider here.  First, in the above an elastic material medium that was surrounded by a like medium on all sides was considered---does a similar argument hold when these conditions do not hold?  As an alternative example consider surface volume elements made up of a noncompressible fluid.  As a realistic concrete example consider the ocean surface where a standard normal reference ellipsoid model has been subtracted off.  Consider the following three well known facts \cite{HandM}:  (1) The sea surface is a surface of constant potential.  (2) The difference in altitude between this surface displacement and the normal ellipsoid is called geoid height (usually denoted H).  (3) The geoid height is proportional to the difference between the potential at the point in question on the surface and on the normal ellipsoid itself.  Notice that while the density and sides of a surface volume element are fixed, the height varies and this change in height (the geoid height) leads directly to a change in mass that is proportional to the change in geoid height.  (Alternatively for the point mass fitting algorithm the point mass values tend to be proportional to the potential difference that is fit, which is proportional in turn to the geoid height [in performing a point mass surface fit here for the region $\Omega_{\sps}$, one generally chooses an appropriate $R_{\sps} < R_E$ so that the point masses do not tail-off away from $\vec{X}_k'$ either too fast or to slow].)  Thus here the energy is proportional to $|m_k|$ rather than $m_k^2$, so perhaps a better form of regularization would be to take $\alpha\sum_{k=1}^{N_k}m_k^2 + \beta \sum_{k=1}^{N_k}|m_k|$ as a regularization factor.  Perhaps even a factor proportional to $\sum_{k=1}^{N_k}|m_k|^{\mu}$, where $1 < \mu < 2$, should be considered.  Here the inclusion of a regularization term with $1 \leqq \mu < 2$ leads to a nonlinear equation set for the $m_k$'s, which is clearly inconvenient.  Furthermore, since minimization of the form $\bar{\Omega}_1$ tends to minimize $\sum_{k=1}^{N_k}|m_k|^{\mu}$ as well, only the form  $\bar{\Omega}_1$ will be considered in the sequel (but this is clearly one of many numerous open issues).   To summarize, when a linear restorative force is present (as one might expect for material stresses and strains) there is a well known quadratic dependence of (potential) energy, but when the displacement mechanism is directly related to the action of gravity on fluid surfaces, the potential energy tends to go like the well known $m\,g\,H$ factor encountered in elementary physics books.

  The final two of the three objections to $\bar{\Omega}_1$ can be stated briefly.  The second one is that $\bar{\Omega}_1$ tends to minimize the overall point mass values and if the subtracted reference model is not as detailed as required this will lead directly to systematic errors in the estimates.  Thirdly and perhaps more importantly: Does even a first order approximation to the required reference model exist for cases of interest?   These last two issues lead to a consideration of other regularization forms.  There is a clear hint in the analysis performed above that leads to an improvement.  In particular, since the stress arising on block $k$ probably originated from a neighboring block in contact with it, a better model to consider is perhaps a regular grid of coupled blocks that can be viewed as a three dimensional assembly of masses that are coupled by springs in the vertical and horizonal directions.  When such a system is in homogenous static equilibrium, the distortional energy is zero, but when each block is either compressed or stretched the resulting total energy will increase.  Thus when one mass is displaced due to pressure from an adjacent mass, not only will the density of that particular block be increased, but also the energy of the block that is directly coupled to it.  The configuration energy of such a coupled pair (just as for a coupled string configuration) can thus be represented by $\omega_{k,\,k'} (m_k - m_{k'})^2$.  By minimizing this coupling energy, a smooth density profile results and the effect of the mass reduction effect is not as pronounced as it is with the straightforward regularization term $\bar{\Omega}_1$.  For ease of implementation here a quadratic form is desirable and it is also necessary to try to enforce (or at least strongly encourage) $m_k \approx m_{k'}$ for $|\vec{X}'_k - \vec{X}'_{k'}| \approx 0$.  A regularization form that fulfills these requirements is
\begin{equation}\label{E:reg2}
\bar{\Omega} = \bar{\Omega}_2 \eq \sum_{k=1}^{N_k}\sum_{k'=1}^{N_k}\omega_{k,\,k'} (m_k - m_{k'})^2
\end{equation}
where $\omega_{k,\,k'}$ produces mass correlation effects.  Thus, in general $\omega_{k,\,k'} = \omega_{k,\,k'} = \omega(d_{k,\,k'})$ with $d_{k,\,k'} := |\vec{X}'_k - \vec{X}'_{k'}|$.  Here, in particular, one suitable choice might be $\omega(d) = 1$ if $d$ is less than $\sqrt{3}$ times the (three-dimensional) grid spacing and $\omega(d) = 0$ otherwise so that only the closest neighbors are correlated. 
This choice of regularization function reduces the condition number and introduces a uniformity into the $m_k$ values without reducing the overall mass values.  This general form of regularization is desirable for many unrelated applications as well.  With the right choice of $\omega$ and $\alpha$, one can clearly negotiate very small grid spacing.

 Before considering several generally desirable refinements to this regularization process, notice that if one assumes that the variations in point mass values is proportional to the underlying local mass deficits or excesses, and that these excesses and deficits are, in turn, proportional to an energy shift due to a local dislocation in the underlying material medium (which, by a standard Taylor's series argument, represents the displacement from what would otherwise have been a local energy minimum), then $\bar{\Omega}_2$ is proportional to energy dislocation of the underlying material medium.  This point will be elaborated on below.  Mathematically $(m_k - m_{k'})^2 = m_k^2 + m_{k'}^2 - 2m_k\,m_{k'}$, so $\bar{\Omega}_2$ contains quadratic terms that effectively add to the diagonal of $T_{k,\,k'}$ as well a bilinear terms, which effectively subtract from the larger off-diagonal elements of $T_{k,\,k'}$ (when $|\vec{X}'_k - \vec{X}'_{k'}| \approx 0$ $T_{k,\,k'} \approx 1$).  Thus $\bar{\Omega}_2$ actually has a stronger regularizing effect than $\bar{\Omega}_1$ (all other things being equal).  

  The foregoing regularization description might best be alternatively encapsulated in terms of a standard Taylor's series argument.  Since $\Phi = \Phi(\mathbf{m})$ (where $\mathbf{m}$ is the vector of mass values), and in particular $\bar{\Omega} = \bar{\Omega}(\mathbf{m})$, it is natural to consider the energy variation in terms of 
$\mathbf{m}$: 
$$\bar{\Omega}(\mathbf{m}_0 + \mathbf{\delta m}) = e^{\,\mathbf{\delta m}\cdot \mathbf{{\nabla}_m}}\bar{\Omega} = \bar{\Omega}({\mathbf{m}}_0) + \mathbf{\delta m}\cdot \mathbf{{\nabla}_m}\,\bar{\Omega}\, + \sum\limits_k\sum\limits_{k'} \delta m_k \delta m_{k'} \frac{{\partial}^2\, \bar{\Omega}\ \ \ \ \ \ }{{\partial m_k}{\partial m_{k'}}} + \cdots$$ 
where $\mathbf{{\nabla}_m}$ has components given by $\frac{\partial\,\ \ }{\partial m_k}$.
In the simplest standard context that this Taylor series argument is used by physicists, an energy minimum, $\mathscr{E}$, is sought and the displacement, $x$, is the variable of interest.  Since it is argued that a physical minimum is present, linear terms cannot be present so the form
 $$ \mathscr{E}(x) = \mathscr{E}_0 + x^2\mathscr{E}'' + \cdots $$
results (which ignores the possibility of physical terms of the form $|x|$).  Since a minimum of $\Phi$ is sought, the constant term for $\bar{\Omega}(\mathbf{m})$ can be ignored, and if it is assumed that a suitable reference has been subtracted off, $\mathbf{m}$ can be identified with $\mathbf{\delta m}$; nevertheless, the linear terms obviously can not be ignored here.  There are three reasons for this.  First, $\bar{\Omega}$ is part of $\Phi$ and linear terms definitely cannot be ignored in the rest of $\Phi$ [see (\ref{E:llsqeqn}), for example] since the linear terms might cancel out in some fashion.  Second, the regularizing functions discussed below (i.e., $\sum [M_0 - m_k]^2$) have linear factors as well as bi-linear terms ($\sum [m_k - m_{k'}]^2$); moreover, these regularization forms have been shown to posses physical relevance.  Third, as noted above, regularization factors proportional to $\sum |m_k|$ have a reasonable physical basis.  At any rate, the physical significance of the first few Taylor series terms should be apparent.  Finally note that, with respect to incompressible fluids and/or stratified media, the above regularization analysis is incomplete at best.  Here the most relevant factor is the shape of the media boundary surface separating one density layer type form another (consider, for example, the ocean floor).  (Hopefully some sort of future analysis undertaken by others will demonstrate a more refined understanding of the higher order aspects and of appropriate regularization functions, in general.)  A means of meshing regularization and surface boundary information will thus be considered next.

  Suppose there are various regions, or density layers, which are distinct, but that each such region, or layer, tends to be homogenous.  This situation can easily be modeled by using a proper choice of $\omega_{k,\,k'}$ in $\bar{\Omega}_2$.  Thus let $\{\mathcal{R}_J\}_{J=1}^{N_J}$ for $J = 1,\, 2,\, 3,\, \ldots,\, N_J$ be a suitable partition of $\Omega_s'$ into subregions: 
$$\bigcup_{J=1}^{N_J}\mathcal{R}_J = \Omega_s'\ \ \text{and}\ \ \mathcal{R}_J\cap\mathcal{R}_{J'} = \emptyset\ \ \text{for}\ J \neq J'$$ 
where $J' = 1,\, 2,\, 3,\, \ldots,\, N_J$.  Then let
\begin{equation}\label{E:wij}
 \omega_{k,\,k'} = \omega(d_{k,\,k'})\ \ \text{if}\ \ \vec{X}_k'\ \text{and}\ \vec{X}_{k'}' \in \mathcal{R}_J,\ \text{and otherwise let}\ \ \omega_{k,\,k'} = 0.
\end{equation}
  The resulting regularization approach characterized by (\ref{E:wij}) tends to produce independent homogenous densities for each of the separate regions. 

 Alternatively, suppose that rough density profile information is available from seismology (or some other means) and this profile is specified by $\hat{\rho}(\vec{X}')$, then this profile can be discretized:  $\hat{\rho}(\vec{X}') \implies \hat{m}_k$.  Then a regularization term of the form
$$\bar{\Omega} = \bar{\Omega}_3 \eq \sum_{k=1}^{N_k} (m_k - \hat{m}_k)^2$$
may be appropriate.  One special case is $\hat{m}_k = \hat{M}_J$ for $\vec{X}' \in \mathcal{R}_J$, where $\hat{M}_J$ is a constant.  This general type of regularization is clearly appropriate when density variations are highly depth dependent and overall density averages are known for certain depths, such as the core or deep mantle.


\SubSec {Structured Point Source Technique (SPST) Basis Functions}

  This subsection first briefly recaps some of the points made in Section~(ii), where the general ideal of a SPST basis function was introduced, and then gives a mathematical description.

  Thus, as previously noted, a common gravitational source density prediction strategy is to divide up the source region into a collection of regular homogenous bodies---with the most common example being to divide up $\Omega_s'$ into arrangement of (nonintersecting) regular parallelepipeds that come close to covering the entire region of interest.  This has an intrinsic regularizing effect.   A very flexible way to implement such a scheme in the present formalism involves approximating the field of each such subregion by a regular grid of closely spaced point masses.  When a collection of point masses share a single common constant mass value (or have fixed relative mass values) and thus have only one undetermined source term, the resulting structure will be labeled a structured point source.  In order to better understand this from a regularization perspective, it is useful to compare this to some of the regularization schemes just considered.  Clearly $\bar{\Omega}_2$ generally has the effect of forcing $m_k \approx m_{k'}$ for nearby masses $m_k$ and $m_{k'}$; moreover, when it is implemented according to the partitioned region regularization scheme, as characterized by using (\ref{E:wij}), similar end results to a SPST basis function fit might well be expected (although one might reasonably argue that SPST basis functions have a stronger regularization effect).  Next consider the regularization effects of $\bar{\Omega}_3$ versus SPST basis functions.  Again one might expect similar end results for most implementations, but clearly, when all else is equal, the automatic constraint implicit in the structured point source technique will have a stronger regularizing effect. 

 As above, let $\{\mathcal{R}_J\}_{J=1}^{N_J}$ be a partition of $\Omega_s'$ into subregions: 
$$\bigcup_{J=1}^{N_J}\mathcal{R}_J = \Omega_s'\ \ \text{and}\ \ \mathcal{R}_J\cap\mathcal{R}_{J'} = \emptyset\ \ \text{for}\ J \neq J'$$ 
where $J = 1,\, 2,\, 3,\, \ldots,\, N_J$ and $J' = 1,\, 2,\, 3,\, \ldots,\, N_J$.
Then the basic SPST idea is to hold the mass fixed over each subregion: $m_k \eq \mathscr{M}_J$ for all $k$ such that $\vec{X}'_k \in \mathcal{R}_J$.  One advantage of this approach over the regularization approach described by (\ref{E:reg2}) and (\ref{E:wij}) is that the resulting linear equation set has dimension $N_J \times N_j$ rather than $N_k \times N_k$.  In detail, let $j$ be a local index for each of the $\mathcal{R}_J$ and let $n(J)$ be the number of (uniform) point sources in $\mathcal{R}_J$, then $\sum_{J=1}^{N_J} n(J) = N_k$ and $1 \leqq j \leqq n(J)$.  Further let $K(J,\,j)$ denote a reordering of the index $k$ such that for all $1 \leqq j \leqq n(J)$ and $1 \leqq J \leqq N_J$
$$ \vec{X}'_K(J,\,j) \in \mathcal{R}_J\ .$$  Then the resulting point mass potential field can be written
$$ V_{N_k}\, = \negthickspace\negthickspace\mLarge{\sum\limits_{\mSmall{{J=1}}}^{\mSmall{{N_J}}} }\negthinspace\negthickspace \mathscr{M}_J \negthickspace\negthickspace \mLarge{\sum\limits_{\mSmall{j=1}}^{\mSmall{n(J)}} }\negthinspace \frac 1{\big|\vec{X} - \vec{X}'_{K(J,\,j)}\big|}$$
and the resulting DIDACKS SPST linear equation set for the $\mathscr{M}_J$'s is
$$\sum\limits_{J'=1}^{N_J} T_{J,\,J'}\,\mathscr{M}_{J'} = A_{J}\,.$$
The required expressions for the half-space energy norm are
$$A_J = \frac14 \sum\limits_{j=1}^{n(J)}W\big(\vec{P}_{K(J,\,j)}\big)$$
\vskip -.15in

\noindent 
and

\vskip -.25in
$$T_{J,\,J'}\, = \, \frac1{4}\negthinspace\negthickspace\negthickspace \mLarge{\sum\limits_{\mSmall{j=1}}^{\mSmall{n(J)}}}\negthinspace\negthickspace\negthickspace\negthickspace\mLarge{\sum\limits_{\mSmall{j'=1}}^{\mSmall{n(J')}} }\frac1{\big|\vec{P}_{K(J,\,j)} - \vec{X}'_{K(J',\,j')}\big|}\,\,.\ \ \ {\text{Here}}\ \ \vec{P}_{K(J,\,j)} = \big( x'_{K(J,\,j)},\,y'_{K(J,\,j)},\,-z'_{K(J,\,j)}\big)^T\ .$$

  Observe that the above description of SPST basis functions can be easily generalized along the lines discussed in Section~(ii) by introducing fixed relative scaling factors.  The idea, of course, is to fix the relative ratios of the various component point masses and then determine the overall scale of the configuration by the fitting process.  Hence, for a set of point masses $m_k$, using the given rations $c_k$, set $m_k = c_k\mathcal{M}_J$, for $\vec{X}_k' \in \mathcal{R}_J$, where only the $M_J$ need to be determined and these values can be determined much the same as before.  This idea has a number of potentially useful applications along the lines of the examples mentioned at the end of Section~(ii).  Some implementation points, however, may not be completely transparent. Suppose, for example, that the boundary of the regions $\mathcal{R}_J$'s are not known and that one wants to estimate these boundaries by a means of DIDACKS NLLSQ scheme.  It is easy enough to set up the problem and get an appropriate cost function to use, but there is an underlying fixed grid for the point sources locations, so that when the surface moves a source point will abruptly switch from one region to another.  A NLLSQ algorithm generally requires good partial derivative information to work well (although there are specialized discrete optimization approaches) so there is a problem.  Thus the problem is that the point mass grid spacing will generally be such that a perturbation of the boundary may not change the cost function, or it may change abruptly all at once, so that the resulting partial derivative information will not be acceptable for NLLSQ purposes.  The easiest solution to this problem is to use the above idea involving point mass ratios in order to implement ``soft boundaries.''  This is easily accomplished as follows. Although the details may not be of interest to many readers, they are included here since they show some of the power and flexibility of the SPST basis function concept.

 For concreteness suppose that the half-plane region $\Omega_1$ is of interest and that the boundaries of $\mathcal{R}_J$ are generally  parallel to the $x'$, $y'$ and $z'$ axes.  In this case it is especially easy to parameterize the boundary surfaces of $\mathcal{R}_J$.  Observe that, because of shared boundaries, when $\mathcal{R}_J$ is surrounded on all sides one can, for accounting purposes, assume that three of the six sides are associated with $\mathcal{R}_J$, so that there are nominally six values of $\{{\eta}_i\}$ associated with each $\mathcal{R}_J$. 

 The overall point is to keep the idea of a well delineated boundary and to keep a fixed SPST  basis function for each $\mathcal{R}_J$, but to define the basis functions in such a way as to make a smooth transition in mass profile across the boundary, without leaving gaps---that is, if the given potential field is constant and uniform over a prediction region, then the resulting mass estimates should be too for this case.   For purposes of this definition, let each region  $\mathcal{R}_J$ and its adjacent neighbor  $\mathcal{R}_{J'}$ overlap by a certain distance $D$.  The idea is that over this distance a gradual transition is made from the uniform mass of  $\mathcal{R}_J$ to the uniform mass of $\mathcal{R}_{J'}$.  For example, if $\mathcal{R}_J$ and its adjacent neighbor $\mathcal{R}_{J'}$ share a common face along the $x-$axis then the $x$ coordinates for the transition might be labeled $X_{T}' \leqq x' \leqq X_T' + D$ where $X_T' = \text{constant} + \eta_i$ for some $i$ (and the constant here is taken so that the midpoint of this interval corresponds to the boundary of $\mathcal{R}_J$ and its adjacent neighbor $\mathcal{R}_{J'}$).  Then if $\vec{X}_k$ is a point in this transition zone, when $\vec{X}_k'$ is treated as a point in  $\mathcal{R}_J$ one should set the corresponding value of the mass for this region to $h(x_k' - X_T')M_J$ where $h(x)$ is a fifty percent cosine taper or $h(x) = x/D$.  [A fifty percent cosine taper for the interval $[o,\,2\pi]$ is simply $ (1 +\cos x)/2$.]  When the same point is considered as a point in $\mathcal{R}_{J'}$ then the corresponding mass value is set to $[1 - h(x_k' - X_T')]M_J'$ so that there is a gradual transition from $\mathcal{R}_J$ to $\mathcal{R}_J'$.  When the DIDACKS cost function with these SPST basis functions is set up, it will have the form $\Phi(\mathcal{M}_J\,\mathbf{\eta})$, where $\mathbf{\eta}$ is the vector of the entire set of surface parameters.  Minimizing this cost function for both $\mathcal{M}_J$ and the components of ${\mathbf{\eta}}$ requires a NLLSQ algorithm for its solution.

  Next consider the second application of the above mass ratio concept.  Although the following idea is general, for the sake of simplicity consider the case where there is only one region of interest so that $\mathcal{R}_1 \eq \Omega_s'$ and 
$$V_{N_k}(\vec{X})\,\, = \negmedspace\negthickspace\negthickspace\mLarge{\sum\limits_{\mSmall{k=1}}^{\mSmall{N_k}} }\negmedspace\! \frac{m_k}{|\vec{X} - \vec{X}_k'|}\ .$$
The central idea here is that instead of using $m_k$ directly as a fitting parameter,  new set of mass fitting parameters is introduced:
$$m_k = m_k(C_i) = \sum\limits_{i=1}^{N_C} C_i \Psi_i({\vec{X}_k'})$$
where the $\Psi_i$'s are a set of $N_C$ suitable basis functions.  The minimization of the resulting cost function $\Phi$ yields a linear set of $N_C$ equations for the $C_i$'s.  Generally here one might place the $\vec{X}_k'$ on a (tight) three dimensional uniform grid; however, it is also possible to arrange the $\vec{X}_k'$'s on a surface grid or along a line array.  When the $\vec{X}_k$'s are arranged on a line a good choice for the $\Psi_i$'s might be a set of Fourier series basis functions expressed as functions of path length along the line.  In these sort of approaches regularization should still be applied as needed.  Also observe that one can easily extend this idea to separate basis expansions over each of the subregions $\mathcal{R}_J$, so no real loss of generality resulted from considering the special case $\mathcal{R}_1 \eq \Omega_s'$.  Finally, the actual form of the linear equation sets that result for these discrete parameterized fits are easily written down and the implementation details are straightforward.  Moreover, the resulting equations are quite similar to those that result from parameterized continuous distributions, which we now turn to (where integrals just basically replace sums).

\SubSec {Continuous Source Estimation}

  It is a small step from the discrete parameterized fits just considered to the consideration of full continuous distributions.  Continuous distributions require numerical integration, so they are more difficult to implement.  To streamline the presentation only the $\Omega_1$ case will be considered here (the $\Omega_{\sps}$ case follows in a like fashion).   Here 
 $$V(\vec{X})\,\, = \negthickspace\mLarge{\iiint\limits_{\Omega_s'}}\negthickspace\negthickspace \frac{\rho{\ls}_V(\vec{X}')}{|\vec{X} - \vec{X}'|}\,\,d^3\,\vec{X}'$$
where $\rho_V$ is a parameterized continuous distribution, which can be taken to have the following form
$$\rho{\ls}_V(\vec{X}') = \sum\limits_{n=1}^{N_C} C_n{\psi}_n(\vec{X}')\ .$$
Here again, the $\Psi_n$'s are a set of suitable basis functions.  The resulting linear equation set for the $C_n$'s is 
           $$\sum\limits_{n=1}^{N_C} T_{n',\,n}C_n = A_n$$
where 
$$T_{n',\,n} = \frac14\mLarge{\iiint\limits_{\Omega_s'}} \mLarge{\iiint\limits_{\Omega_s'}} \frac{\Psi(x',\,y',\,-z')\Psi(x'',\,y'',\,-z'')}{\sqrt{(x' -x'')^2 + (y' - y'')^2 + (z' + z'')^2}}\, d^3\,\vec{X}'\,d^3\,\vec{X}''$$
with $\vec{X}'$ and $\vec{X}'' \in \Omega_{s}'$.  Likewise
$$ A_n = \frac14\iiint\limits_{\Omega_s'}\ {W(x',\,y',\,-z')\Psi(x',\,y',\,-z')}\, d^3\,\vec{X}'\ .$$



\begin{thebibliography}{99}
\bibitem{OldRef}
C. A. Heiland,
\emph{Geophysical Exploration},
Hafner Publishing Co., New York and London, 1968;
reprint of 1940 Prentice-Hall edition.
\bibitem{HandM}
Bernhard Hofmann-Wellenhof and Helmut Moritz,
\emph{Physical Geodesy},
Springer-Verlag New York, New York, 2005.
\bibitem{Jard}
Wenceslas S. Jardetzky,
\emph{Theories of figures of Celestial Bodies},
Dover Publications, New York, 1995 \{Reprint of 1958 edition\}.
\bibitem{Kellogg}
Oliver Dimon Kellogg,
\emph{Foundations of Potential Theory},
Dover Publications, New York, 1953;
reprint of 1929 edition.
\bibitem{LandH1}
C. Lawson and R. Hanson,
\emph{Solving Least Squares Problems},
 First Edition,
Prentice--Hall, Englewood Cliffs, N. J., 1974.
\bibitem{Moritz}
Helmut Moritz,
\emph{Advanced Physical Geodesy},
Abacus Press, Tunbridge Wells, Kent, England, 1980.
\bibitem{AppliedGeodynamics}
Jiri Nedoma,
\emph{Numerical Modeling in Applied Geodynamics},
John Wiley \& Sons, Inc., New York, 1998.
\bibitem{Parker}
Robert L. Parker,
\emph{Geophysical Inverse Theory},
Princeton University Press, Princeton, New Jersey, 1994.
\bibitem{LDP}
Alan Rufty,
\emph{Comments on the Reliability of Lawson and Hanson's Linear Distance Programming Algorithm: Subroutine LDP},
[arxiv:0707.9651].
\bibitem{DIDACKS}
Alan Rufty,
\emph{A Dirichlet-Integral Based Dual-Access Collocation-Kernel Approach to Point-Source Gravity-Field Modeling}, SIAM Journal on Applied Mathematics, \textbf{68}, No. 1, 199--221.
\bibitem{DIDACKSI}
Alan Rufty,
\emph{Dirichlet integral dual-access collocation-kernel space analytic interpolation for unit disks: DIDACKS I}, [arxiv:math-ph/0702062].
\bibitem{DIDACKSII}
Alan Rufty,
\emph{Dirichlet-integral point-source harmonic interpolation over ${\mathbb{R}}^3$ spherical interiors: DIDACKS II}, [arxiv:math-ph/0702063].
\bibitem{DIDACKSIII}
Alan Rufty,
\emph{Closed-form Dirichlet integral harmonic interpolation-fits for real n-dimensional and complex half-space: DIDACKS III}, [arxiv:math-ph/0702064].
\bibitem{Tarantola}
A. Tarantola,
\emph{Inverse Problems = Quest for Information}, Journal of Geophysics, \textbf{50}, 159--170.
\bibitem{Tarantola2}
Albert Tarantola,
\emph{Inverse Problem Theory, Methods for Data Fitting and Model Parameter Estimation},
 Elsevier Science Publishers, 1987.
\bibitem{Todhunter}
I. Todhunter,
\emph{A History of the Mathematical Theories of Attraction and the Figure of the Earth, from the Time of Newton to that of Laplace},
In Two Volumes, MacMillan and Co., London, 1873.
\bibitem{Geodynamics}
Donald L. Turcotte and Gerald Scubert,
\emph{Geodynamics},
Second Edition,
Cambridge University Press, New York, N.Y., 2001.
\end{thebibliography}
\end{document}